\documentclass[12pt]{article}
\usepackage{epsfig,amsfonts,amsthm}
\usepackage{color}
\usepackage{amsmath,amssymb}
\usepackage{array}
\newcommand{\be}{\begin{equation}}
\newcommand{\ee}{\end{equation}}
\newcommand{\bea}{\begin{eqnarray}}
\newcommand{\eea}{\end{eqnarray}}

\renewcommand{\Re}{\mathrm{Re }}
\renewcommand{\Im}{\mathrm{Im }}
\newcommand{\doublet}[2]{ \left( \begin{array}{c}#1 \\ #2 \end{array}\right) }

\def\lsim{\mathrel{\rlap{\lower4pt\hbox{\hskip1pt$\sim$}}
    \raise1pt\hbox{$<$}}}         %less than or approx. symbol
\def\gsim{\mathrel{\rlap{\lower4pt\hbox{\hskip1pt$\sim$}}
    \raise1pt\hbox{$>$}}}         %greater than or approx. symbol

\usepackage{amssymb}
\usepackage{subfig}
\usepackage{wrapfig}
\usepackage{graphicx}

\def\beq{\begin{equation}}
\def\eeq{\end{equation}}
\def\bea{\begin{eqnarray}}
\def\eea{\end{eqnarray}}

\def\<{\left\langle}
\def\>{\right\rangle}

\newcommand{\bt}{\begin{tabular}}
\newcommand{\et}{\end{tabular}}

\allowdisplaybreaks[2]
\begin{document}
\bibliographystyle{OurBibTeX}

\title{\hfill ~\\[-40mm]
                  \textbf{\Large Phenomenology of the Inert (2+1) and (4+2) \\
                  Higgs Doublet Models
                  %Z$_2$-symmetric 3HDM:\\ a multi-inert-doublet model
                }        }
%\date{}
\author{\\[-7.5mm]
Venus~Keus\footnote{E-mail: {\tt Venus.Keus@helsinki.fi}} $^{1,2,3,4}$,\ 
Stephen F. King\footnote{E-mail: {\tt King@soton.ac.uk}} $^{2}$,\ 
Stefano~Moretti\footnote{E-mail: {\tt S.Moretti@soton.ac.uk}} $^{2,4}$ 
\\[-1mm]
  \emph{\small $^1$ Department of Physics and Helsinki Institute of Physics,}\\
  \emph{\small Gustaf Hallstromin katu 2, FIN-00014 University of Helsinki, Finland}\\
  \emph{\small $^2$ School of Physics and Astronomy, University of Southampton,}\\
  \emph{\small Southampton, SO17 1BJ, United Kingdom}\\
  \emph{\small  $^3$ Department of Physics, Royal Holloway, University of London,}\\
  \emph{\small Egham Hill, Egham TW20 0EX, United Kingdom}\\
  \emph{\small $^4$ Particle Physics Department, Rutherford Appleton Laboratory,}\\
  \emph{\small Chilton, Didcot, Oxon OX11 0QX, United Kingdom}\\[0mm]}

\maketitle

\begin{abstract}
\noindent
We make a phenomenological study of a model with two inert doublets 
plus one Higgs doublet (I(2+1)HDM)
which is symmetric under a Z$_2$ group, preserved after Electro-Weak Symmetry Breaking (EWSB) by the vacuum alignment $(0,0,v)$. This model may be regarded as an extension to the model with one inert doublet plus one Higgs doublet (I(1+1)HDM), by the addition of an extra inert scalar doublet. The neutral fields from the two inert doublets provide a viable Dark Matter (DM) candidate which is stabilised by the conserved $Z_2$ symmetry. We study the new Higgs decay channels offered by the scalar fields from the extra doublets and their effect on the Standard Model (SM) Higgs couplings, including a new
decay channel into (off-shell) photon(s) plus missing energy, which distinguishes the I(2+1)HDM from the
I(1+1)HDM. Motivated by Supersymmetry, which requires an even number of doublets, we then extend this model into a model with four inert doublets plus two Higgs doublets (I(4+2)HDM) and study the phenomenology of the model with the vacuum alignment $(0,0,0,0,v,v)$. This scenario offers a wealth of Higgs signals, the most distinctive ones being cascade decays of heavy Higgs states into inert ones. Finally, we also remark that the smoking-gun signature of all the considered models is represented by   invisible Higgs decays into the lightest inert Higgs bosons responsible for DM.
%for $\tan\beta=1$ case and the general $\tan\beta \neq 1$ case.
\end{abstract}
\thispagestyle{empty}
\vfill
\newpage
\setcounter{page}{1}

\section{Introduction}
The Standard Model (SM) is the original one-Higgs-doublet model (1HDM) and represents the minimal model of Electro-Weak Symmetry Breaking (EWSB). Although the observation of one Higgs boson, with properties that are
consistent with those predicted by the SM, lends support to the 1HDM, it is entirely possible that further Higgs bosons could be discovered in the next run of the Large Hadron Collider (LHC). Such a discovery would require additional Higgs multiplets.

The simplest example with extra Higgs doublets is the class of 2-Higgs-Doublet Models 
(2HDMs) \cite{Gunion:1989we},
for example the type II model predicted by the Minimal Supersymmetric Standard Model 
(MSSM) \cite{Chung:2003fi}.  Of course, the 
general class of 2HDMs is much richer than this example, and indeed all possible types 
of 2HDMs have been well studied in the literature \cite{Chang:2013ona}. However, general 2HDMs face challenging phenomenological problems with Flavour Changing Neutral Currents (FCNCs) and 
possible charge breaking vacua, and it is common to consider restricted classes of models controlled
by various symmetries. 
The introduction of symmetries into 2HDMs provides a welcome restriction to the
rather unwieldy general Higgs potential, as well as solutions to the phenomenological challenges mentioned above. For example, the remaining symmetry of the potential after EWSB can have the effect of stabilising the lightest Higgs boson, which can become a possible Dark Matter (DM) candidate. 
In 2HDMs, the full list of possible 
symmetries of the potential is now known \cite{2HDM}.

There are good motivations for considering the 
case of 3-Higgs-Doublet Models (3HDMs). 
To begin with, it is the next simplest example beyond 2HDMs, which have been extensively studied in the
literature. Furthermore, for 3HDMs, all possible finite symmetries have been identified \cite{Ivanov:2011ae, Ivanov:2012fp}.
Intriguingly, 3HDMs may shed light on the flavour problem, namely the problem
of the origin and nature of the three families of quarks and leptons, including neutrinos, and their 
pattern of masses, mixings and CP violation. Indeed
it is possible that the three families of quarks and leptons could be described by the same symmetries that describe the three Higgs doublets. 
In such models this family symmetry could be spontaneously broken along with the EW  symmetry, although some remnant subgroup could survive, thereby stabilising a possible scalar DM candidate. 
For certain symmetries it is possible to find a Vacuum Expectation Value (VEV) alignment that respects the original symmetry of the potential which will then be responsible for the stabilization of the DM candidate.

In a recent paper \cite{Keus:2013hya}, we discussed the classification of 3HDMs in terms of all possible Abelian symmetries (continuous and discrete) and all possible discrete non-Abelian symmetries. 
We analysed the potential in each case and derived the conditions under which 
the vacuum alignments $(0,0,v)$, $(0,v,v)$ and $(v,v,v)$ are minima of the potential.
For the alignment $(0,0,v)$, which is of particular interest because of its relevance for DM models and the absence of FCNCs, we calculated the corresponding 
physical Higgs boson mass spectrum.
This led to phenomenological constraints on the parameters
in the potential and, for certain parameter choices, the possibility of additional
light Higgs bosons which may have evaded detection at the Large Electron-Positron (LEP) collider. 
It is possible that the 125 GeV Higgs boson could decay
into these lighter Higgs bosons, providing striking new signatures for Higgs decays at the LHC. 
Motivated by Suspersymmetry, we also extended the analysis
to the case of three up-type Higgs doublets and three down-type Higgs doublets (six doublets in total).

In the present paper we shall focus on the phenomenology of one of these 3HDMs, namely the one
based on the minimal Abelian symmetry $Z_2$ under which the third Higgs doublet is even and the first and second Higgs doublets are odd. The $Z_2$ symmetry is therefore preserved for the vacuum alignment 
$(0,0,v)$ which we assume in this paper. Thus we are led to consider a a model with two inert doublets plus one Higgs doublet (I(2+1)HDM). This model may be regarded as an extension of the model with one inert doublet plus one Higgs doublet (I(1+1)HDM)\footnote{This model is known in the literature as the Inert-doublet model (IDM). We refer to this model as (I(1+1)HDM) for the clarification of the number of inert and active Higgs doublets.} proposed in 1976 \cite{Deshpande:1977rw} and has been studied extensively for the last few years (see, e.g., \cite{Ma:2006km,Barbieri:2006dq,LopezHonorez:2006gr}), by the addition of an extra inert scalar doublet. The doublet is termed ``inert'' since it does not develop a
VEV, nor does it couple to fermions.
The lightest neutral field from the two inert doublets provides a 
viable DM candidate which is stabilised by the conserved $Z_2$ symmetry. We study the new Higgs decay channels offered by the scalar fields from the extra doublets and their effect on the SM Higgs couplings. Again,
motivated by Supersymmetry, which requires an even number of doublets,
we then extend this model into a model with four inert doublets plus one Higgs doublet (I(4+2)HDM), for example as in the E$_6$SSM \cite{King:2005jy,King:2005my},
and first study the theory and phenomenology of the Higgs field near the vacuum point $(0,0,0,0,v,v)$ in the special $\tan\beta=1$ case, before turning to the general $\tan\beta \neq 1$ case\footnote{In analogy with 2HDMs, here $\tan\beta$ is the ratio of the VEVs of the two active Higgs doublets.}.

The layout of the remainder of the paper is as follows.
In section \ref{3HDM} we discuss the I(2+1)HDM of interest.
In section \ref{6HDMtb1} we discuss the I(4+2)HDM for the case $\tan\beta=1$.
In section \ref{6HDMtbneq1} we discuss the I(4+2)HDM for the case $\tan\beta \neq 1$.
In each case we write down the potential and its symmetries, before giving the vacuum alignment and the scalar mass spectrum, followed by theoretical and experimental constraints and phenomenological implications. Section \ref{conclusion} concludes the paper.

\section{The I(2+1)HDM}
\label{3HDM}

\subsection{The potential with Z$_2$ symmetry }

In a N-Higgs-Doublet Models (NHDMs), the scalar potential symmetric under a group $G$ of phase rotations can be written as 
\be 
V = V_0 + V_G
\ee
where $V_0$ is invariant under any phase rotation and $V_G$ is a collection of extra terms ensuring the symmetry group $G$ \cite{Ivanov:2011ae}.
The most general phase invariant part of the 3HDM potential has the following form:
\bea
V_0 &=& - \mu^2_{1} (\phi_1^\dagger \phi_1) -\mu^2_2 (\phi_2^\dagger \phi_2) - \mu^2_3(\phi_3^\dagger \phi_3) \\
&&+ \lambda_{11} (\phi_1^\dagger \phi_1)^2+ \lambda_{22} (\phi_2^\dagger \phi_2)^2  + \lambda_{33} (\phi_3^\dagger \phi_3)^2 \nonumber\\
&& + \lambda_{12}  (\phi_1^\dagger \phi_1)(\phi_2^\dagger \phi_2)
 + \lambda_{23}  (\phi_2^\dagger \phi_2)(\phi_3^\dagger \phi_3) + \lambda_{31} (\phi_3^\dagger \phi_3)(\phi_1^\dagger \phi_1) \nonumber\\
&& + \lambda'_{12} (\phi_1^\dagger \phi_2)(\phi_2^\dagger \phi_1) 
 + \lambda'_{23} (\phi_2^\dagger \phi_3)(\phi_3^\dagger \phi_2) + \lambda'_{31} (\phi_3^\dagger \phi_1)(\phi_1^\dagger \phi_3).  \nonumber
\label{V0-3HDM}
\eea
Constructing the $Z_2$-symmetric part of the potential depends on the generator of the group. The $Z_2$ generator which forbids FCNCs and is respected by the vacuum alignment $(0,0,v)$ has the following form
\be 
g=  \mathrm{diag}\left(-1, -1, +1 \right). 
\ee
The terms ensuring the $Z_2$ group generated by this $g$ are
\be 
V_{Z_2} = -\mu^2_{12}(\phi_1^\dagger\phi_2)+  \lambda_{1}(\phi_1^\dagger\phi_2)^2 + \lambda_2(\phi_2^\dagger\phi_3)^2 + \lambda_3(\phi_3^\dagger\phi_1)^2  + h.c. 
\label{Z_2-3HDM}
\ee
which need to be added to $V_0$ in Eq.~(\ref{V0-3HDM}) to result in a potential whose only symmetry is $Z_2$ with no larger accidental symmetry.
%\footnote{Note that removing the $-\mu^2_{12}(\phi_1^\dagger\phi_2)$ term leads to a $Z_2 \times Z_2$ symmetry of the potential generated by $g'=(-1,1,1)$ and $g''=(1,-1,1)$.}

\subsection{Mass eigenstates}

We define the doublets as
\be 
\phi_1= \doublet{$\begin{scriptsize}$ H^+_1 $\end{scriptsize}$}{\frac{H^0_1+iA^0_1}{\sqrt{2}}} ,\quad 
\phi_2= \doublet{$\begin{scriptsize}$ H^+_2 $\end{scriptsize}$}{\frac{H^0_2+iA^0_2}{\sqrt{2}}} , \quad 
\phi_3= \doublet{$\begin{scriptsize}$ H^+_3 $\end{scriptsize}$}{\frac{v+H^0_3+iA^0_3}{\sqrt{2}}}  
\label{explicit-fields}
\ee
with two inert doublets ($\phi_1$ and $\phi_2$) and one active doublet ($\phi_3$).
The CP-even/odd neutral fields from the inert doublets could in principle be DM candidates since only the fields from the active doublet couple to the fermions. To stabilise the DM candidate from decaying into SM particles, we make use of the conserved $Z_2$ symmetry of the potential after EWSB. 
To make sure that the entire Lagrangian and not only the scalar potential is $Z_2$ symmetric, we assign an even $Z_2$ parity to all SM particles, identical to the $Z_2$ parity of the only doublet that couples to them, i.e., the active doublet $\phi_3$. With this parity assignment FCNCs are avoided as the extra doublets are forbidden to decay to fermions by $Z_2$ conservation.

Note that the Yukawa Lagrangian in this model is identical to the SM Yukawa Lagrangian, where $\phi_3$ plays the role of the SM-Higgs doublet:
\bea 
\mathcal{L}_{Yuk} &=& \Gamma^u_{mn} \bar{q}_{m,L} \tilde{\phi}_3 u_{n,R} + \Gamma^d_{mn} \bar{q}_{m,L} \phi_3 d_{n,R} \nonumber\\
&& +  \Gamma^e_{mn} \bar{l}_{m,L} \phi_3 e_{n,R} + \Gamma^{\nu}_{mn} \bar{l}_{m,L} \tilde{\phi}_3 {\nu}_{n,R} + h.c.  
\eea
 
The point $(0,0,\frac{v}{\sqrt{2}})$ becomes the minimum of the potential at 
\be
v^2= \frac{\mu^2_3}{\lambda_{33}} 
\ee

Expanding the potential around this vacuum point results in the mass spectrum below, where the pairs of scalar/pseudo-scalar/charged base fields from  the inert doublets in Eq.~(\ref{explicit-fields}) are rotated by:
\be 
R_{\theta_i}= 
\left( \begin{array}{cc}
\cos \theta_i & \sin \theta_i \\
-\sin \theta_i & \cos \theta_i\\
\end{array} \right), \qquad \theta_i = \theta_h , \theta_a , \theta_c,
\ee
with $\theta_h , \theta_a$ and $\theta_c $ the rotation angles in the scalar, pseudo-scalar and charged mass-squared matrix respectively.
The fields from the third doublet play the role of the SM-Higgs doublet fields.
\begin{footnotesize}
\bea
&& \textbf{G}^0  : \quad m^2_{G^0}=0 \nonumber\\
&& \textbf{G}^\pm  : \quad m^2_{G^\pm}=0 \nonumber\\
&& \textbf{h} : \quad m^2_{h}= 2\mu_3^2 \nonumber\\
&& \textbf{H}_1 = \cos\theta_h H^0_{1}+ \sin\theta_hH^0_{2} :  \quad m^2_{H_1}=  (-\mu^2_1 + \Lambda_{\phi_1})\cos^2\theta_h + (- \mu^2_2 + \Lambda_{\phi_2}) \sin^2\theta_h -2\mu^2_{12} \sin\theta_h \cos\theta_h \nonumber\\
&& \textbf{H}_2 = -\sin\theta_h H^0_{1}+ \cos\theta_hH^0_{2} : \quad m^2_{H_2}=  (-\mu^2_1 + \Lambda_{\phi_1})\sin^2\theta_h + (- \mu^2_2 + \Lambda_{\phi_2}) \cos^2\theta_h + 2\mu^2_{12} \sin\theta_h \cos\theta_h \nonumber\\
&& \qquad \qquad  \mbox{where} \quad \Lambda_{\phi_1}= \frac{1}{2}(\lambda_{31} + \lambda'_{31} +  2\lambda_3)v^2  \nonumber\\
&& \qquad \qquad \qquad \qquad \Lambda_{\phi_2}= \frac{1}{2}(\lambda_{23} + \lambda'_{23} +2\lambda_2 )v^2   \nonumber\\
&& \qquad \qquad \qquad \qquad \tan 2\theta_h = \frac{2\mu^2_{12}}{\mu^2_1 -\Lambda_{\phi_1} - \mu^2_2 + \Lambda_{\phi_2}} \nonumber\\[2mm]
&& \textbf{H}^\pm_1 = \cos\theta_cH^\pm_{1}+ \sin\theta_cH^\pm_{2} : \quad m^2_{H^\pm_1}=  (-\mu^2_1 + \Lambda'_{\phi_1})\cos^2\theta_c + (- \mu^2_2 + \Lambda'_{\phi_2}) \sin^2\theta_c -2\mu^2_{12} \sin\theta_c \cos\theta_c \nonumber\\
&& \textbf{H}^\pm_2 = -\sin\theta_cH^\pm_{1}+ \cos\theta_cH^\pm_{2} :\quad m^2_{H^\pm_2}= (-\mu^2_1 + \Lambda'_{\phi_1})\sin^2\theta_c + (- \mu^2_2 + \Lambda'_{\phi_2}) \cos^2\theta_c + 2\mu^2_{12} \sin\theta_c \cos\theta_c \nonumber\\
&& \qquad \qquad  \mbox{where} \quad \Lambda'_{\phi_1}= \frac{1}{2}(\lambda_{31})v^2  \nonumber\\
&& \qquad \qquad \qquad \qquad  \Lambda'_{\phi_2}= \frac{1}{2}(\lambda_{23} )v^2   \nonumber\\
&& \qquad \qquad \qquad \qquad \tan 2\theta_c = \frac{2\mu^2_{12}}{\mu^2_1 - \Lambda'_{\phi_1} - \mu^2_2 + \Lambda'_{\phi_2}} \nonumber\\[2mm]
&& \textbf{A}_1 = \cos\theta_aA^0_{1}+ \sin\theta_a A^0_{2}:  \quad m^2_{A_1}= (-\mu^2_1 + \Lambda''_{\phi_1})\cos^2\theta_a + (- \mu^2_2 + \Lambda''_{\phi_2}) \sin^2\theta_a -2\mu^2_{12} \sin\theta_a \cos\theta_a \nonumber\\
&& \textbf{A}_2 = -\sin\theta_aA^0_{1}+ \cos\theta_a A^0_{2} : \quad m^2_{A_2}= (-\mu^2_1 + \Lambda''_{\phi_1})\sin^2\theta_a + (- \mu^2_2 + \Lambda''_{\phi_2}) \cos^2\theta_a + 2\mu^2_{12} \sin\theta_a \cos\theta_a \nonumber\\
&& \qquad \qquad  \mbox{where} \quad \Lambda''_{\phi_1}= \frac{1}{2}(\lambda_{31} + \lambda'_{31} - 2\lambda_3)v^2  \nonumber\\
&& \qquad \qquad \qquad \qquad  \Lambda''_{\phi_2}= \frac{1}{2}(\lambda_{23} + \lambda'_{23} -2\lambda_2 )v^2   \nonumber\\
&& \qquad \qquad \qquad \qquad \tan 2\theta_a = \frac{2\mu^2_{12}}{\mu^2_1 - \Lambda''_{\phi_1} - \mu^2_2 + \Lambda''_{\phi_2}} \nonumber
\eea
\end{footnotesize}

\begin{figure}[ht!]
\centering
\includegraphics[scale=0.8]{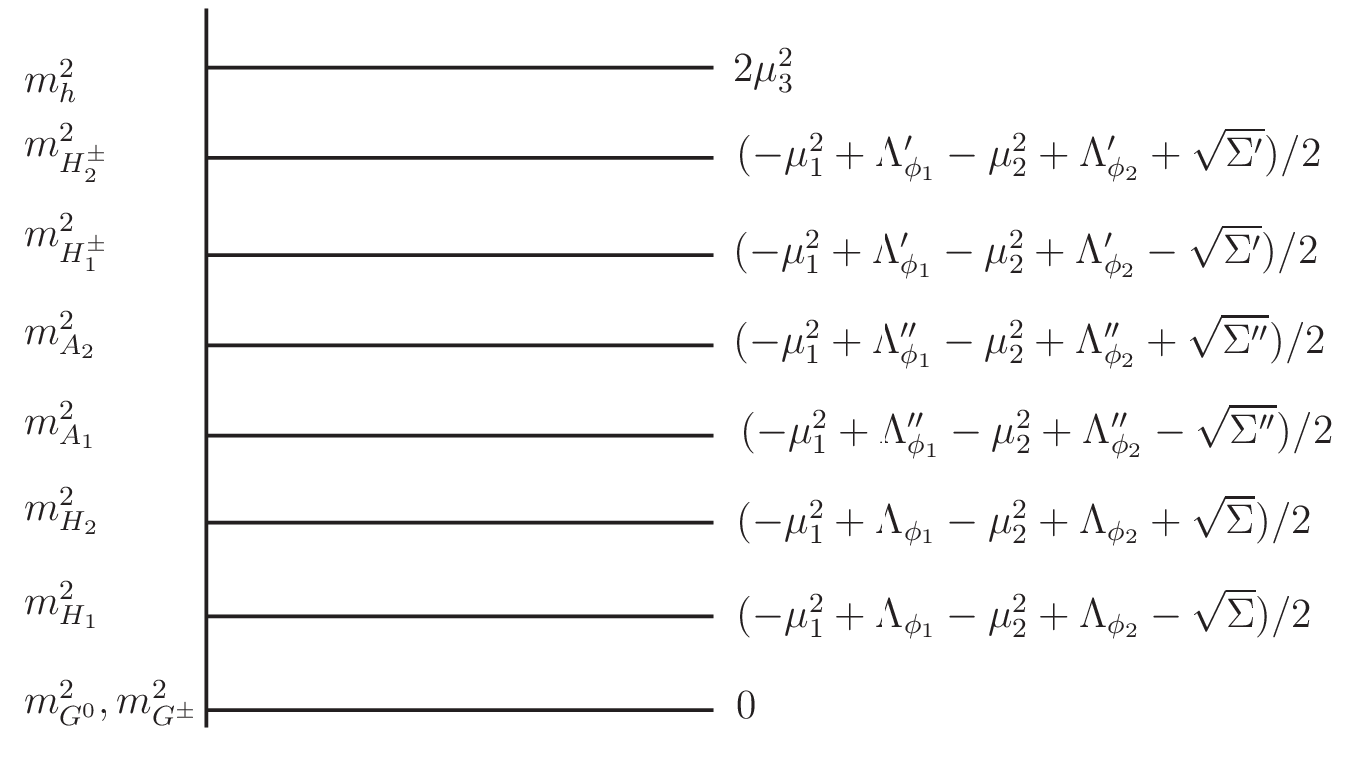} 
\caption{Schematic mass-squared spectrum of the $Z_2$ symmetric 3HDM, where $\Sigma= 4\mu^4_{12} + (\mu^2_1-\Lambda_{\phi_1} -\mu^2_2 +\Lambda_{\phi_2})^2 $, and $\Sigma' = 4\mu^4_{12} + (\mu^2_1-\Lambda'_{\phi_1} -\mu^2_2 +\Lambda'_{\phi_2})^2 $ and $\Sigma''=4\mu^4_{12} + (\mu^2_1-\Lambda''_{\phi_1} -\mu^2_2 +\Lambda''_{\phi_2})^2 $. Note that the third doublet plays the role of the SM Higgs doublet.}
\label{Masses-fig}
\end{figure}

There are two generations of physical inert states: fields from the first generation, $(H_1,A_1,H^\pm_1)$ are chosen to be lighter than the respective fields from the second generation, $(H_2,A_2,H^\pm_2)$, with $H_1$ being the lightest of them all, i.e., the DM candidate:
\begin{equation}
m_{H_1} < m_{H_2}, m_{A_{1,2}},m_{H^\pm_{1,2}}.
\end{equation}
The mass spectrum has the schematic form shown in Fig.(\ref{Masses-fig}), provided the CP-even neutral inert particles are lighter than the CP-odd and charged inert particles, which implies 
\be 
2\lambda_2 , 2\lambda_3 < \lambda'_{23}, \lambda'_{31} < 0. 
\label{lambda-assumption} 
\ee

\subsection{Constraints on parameters}\label{constraints}

The parameters of the potential can be divided into the following categories:
\begin{itemize}
\item $\mu_3, \lambda_{33}$: Higgs field parameters, given by the Higgs mass, 
$$ m^2_h = 2\mu^2_3 = 2\lambda_{33} v^2 $$

\item $\mu_{1},\mu_{2},\mu_{12}, \lambda_{31},\lambda_{23},\lambda'_{31},\lambda'_{23},\lambda_{2}, \lambda_{3}$: masses of inert scalars and their couplings with the visible sector ($h$), these 9 parameters can in principle be determined by 6 masses and 3 mixing angles. 

The ranges we allow for these values in our numerical studies are:
\bea
&& -10 ~\mbox{ TeV}^2 < \mu^2_{1},\mu^2_{2},\mu^2_{12} < 10 ~\mbox {TeV}^2, \nonumber\\
&& -0.5 < \lambda_{31},\lambda_{23},\lambda'_{31},\lambda'_{23},\lambda_{2}, \lambda_{3} < 0.5. \nonumber
\eea

\item $\lambda_{11},\lambda_{22},\lambda_{12}, \lambda'_{12}$: inert parameters (inert scalars self-interactions). Relic density calculations do not depend on these parameters. Any bound on these parameters should therefore come from collider limits.

The ranges we allow for these values in our numerical studies are:
\be 
\bullet \quad 0 < \lambda_{11},\lambda_{22},\lambda_{12}, \lambda'_{12} < 0.5, \nonumber
\ee

\end{itemize}

%\subsection{Theoretical constraints}
Theoretical constraints from positivity of mass eigenstates, bounded-ness of the potential and positive-definite-ness of the Hessian put the following constraints on the potential.
\begin{enumerate}
\item 
\textbf{Positivity of the mass eigenstates}:
\bea
\label{positivity}
&& \bullet \qquad \mu^2_3 > 0    \\
&& \bullet \quad -2\mu^2_1 + \lambda_{31}v^2> 0 \nonumber\\
&& \bullet \quad  -2\mu^2_1 + (\lambda_{31}+\lambda'_{31} )v^2> 0 \nonumber\\
&& \bullet \quad -2\mu^2_1 + (\lambda_{31}+\lambda'_{31} -2\lambda_3 )v^2> 0   \nonumber\\
&& \bullet \quad -2\mu^2_2 + \lambda_{23}v^2> 0 \nonumber\\
&& \bullet \quad -2\mu^2_2 + (\lambda_{23}+\lambda'_{23} )v^2> 0 \nonumber\\
&& \bullet \quad -2\mu^2_2 + (\lambda_{23}+\lambda'_{23} -2\lambda_2 )v^2> 0   \nonumber\\
&& \bullet \quad  -2\mu^2_1 -2\mu^2_2  + (\lambda_{31}+ \lambda_{23})v^2> 4|\mu^2_{12}|   \nonumber\\
&& \bullet \quad -2\mu^2_1 -2\mu^2_2  + (\lambda_{31}+ \lambda_{23} +\lambda'_{31}+ \lambda'_{23} )v^2> 4|\mu^2_{12}|   \nonumber\\
&& \bullet \quad -2\mu^2_1 -2\mu^2_2  + (\lambda_{31}+ \lambda_{23} +\lambda'_{31}+ \lambda'_{23} - 2\lambda_3 -2\lambda_2)v^2> 4|\mu^2_{12}|   \nonumber
\eea

\item
\textbf{Bounded-ness of the potential}:
\\For the $V_0$ part of the potential Eq.~(\ref{V0-3HDM}) to have a stable vacuum (bounded from below) the following conditions are required:
\bea
&& \bullet\quad \lambda_{11}, \lambda_{22}, \lambda_{33} > 0 \\
&& \bullet\quad \lambda_{12} + \lambda'_{12} > -2 \sqrt{\lambda_{11}\lambda_{22}} \nonumber\\
&& \bullet\quad \lambda_{23} + \lambda'_{23} > -2 \sqrt{\lambda_{22}\lambda_{33}} \nonumber\\
&& \bullet\quad \lambda_{31} + \lambda'_{31} > -2 \sqrt{\lambda_{33}\lambda_{11}} \nonumber
\eea
We also require the parameters of the $V_{Z_2}$ part Eq.~(\ref{Z_2-3HDM}) to be smaller than the parameters of the $V_0$ part of the potential:
\be 
\bullet\quad |\lambda_1|, |\lambda_2|, |\lambda_3| < |\lambda_{ii}|, |\lambda_{ij}|, |\lambda'_{ij}| , \quad i\neq j : 1,2,3
\ee

\item
\textbf{Positive-definite-ness of the Hessian}:
\\For the point $(0,0,\frac{v}{\sqrt{2}})$ to be a minimum of the potential, the second order derivative matrix must have positive definite determinant. Therefore, the following constraints are required:
\bea 
&& \bullet\qquad   \mu^2_3  > 0   \\
&& \bullet\quad  -2\mu^2_2  + (\lambda_{23}+\lambda'_{23} )v^2   >0  \nonumber\\
&& \bullet\quad  -2\mu^2_1  + (\lambda_{31}+\lambda'_{31} )v^2   >0  \nonumber\\
&& \bullet\quad  \biggl(-2\mu^2_1  +(\lambda_{31}+\lambda'_{31} )v^2 \biggr) \biggl( -2\mu^2_2  +(\lambda_{23}+\lambda'_{23} )v^2 \biggr) > 4\mu^4_{12}  \nonumber
\eea

\end{enumerate}

\subsection{Results for the I(2+1)HDM }

In this scenario, the only active Higgs boson, $h$, is such that its couplings
to all SM matter are identical to those of the SM Higgs scalar. Therefore, the only effects which could emerge from the
presence of two additional inert Higgs doublets would come from triple Higgs couplings involving the (active) SM state
and two inert objects. This could ultimately result in an increased invisible width of the $h$ state, owing to the possibility of the decays
$h\to {H_{1}}{H_{1}}$, with the ${H_{1}}$ being (by construction) the only
DM candidate of this model.
In order to assess the typical size of this decay, we have performed a scan over the 13-dimensional parameter space of the model, by randomly 
generating points over intervals listed in section ~\ref{constraints}\footnote{Notice that both
$\lambda_{33}$ and $\mu_3$
are fixed in terms of $m_h=125$ GeV.}. In doing so, we keep into account experimental constraints on the lightest neutral
and charged inert Higgs states stemming from LEP data, as follows\footnote{Note that in our Ref.~\cite{Keus:2014jha} we have used a somewhat more restrictive set of experimental constraints, though 
we have verified that no substantial differences emerge in the ensuing phenomenology. Similarly, the I(2+1)HDM parameter space studied here is broadly compliant with the DM restrictions derived therein.}
$$
        m_{H^{\pm}_{1}} > 100~{\rm GeV},
$$
\begin{equation}\label{eq:3HDM-exp_constraints}
         m_{H_{1}} , m_{A_{1}} > m_Z/2.
\end{equation}

\begin{figure}[!t]
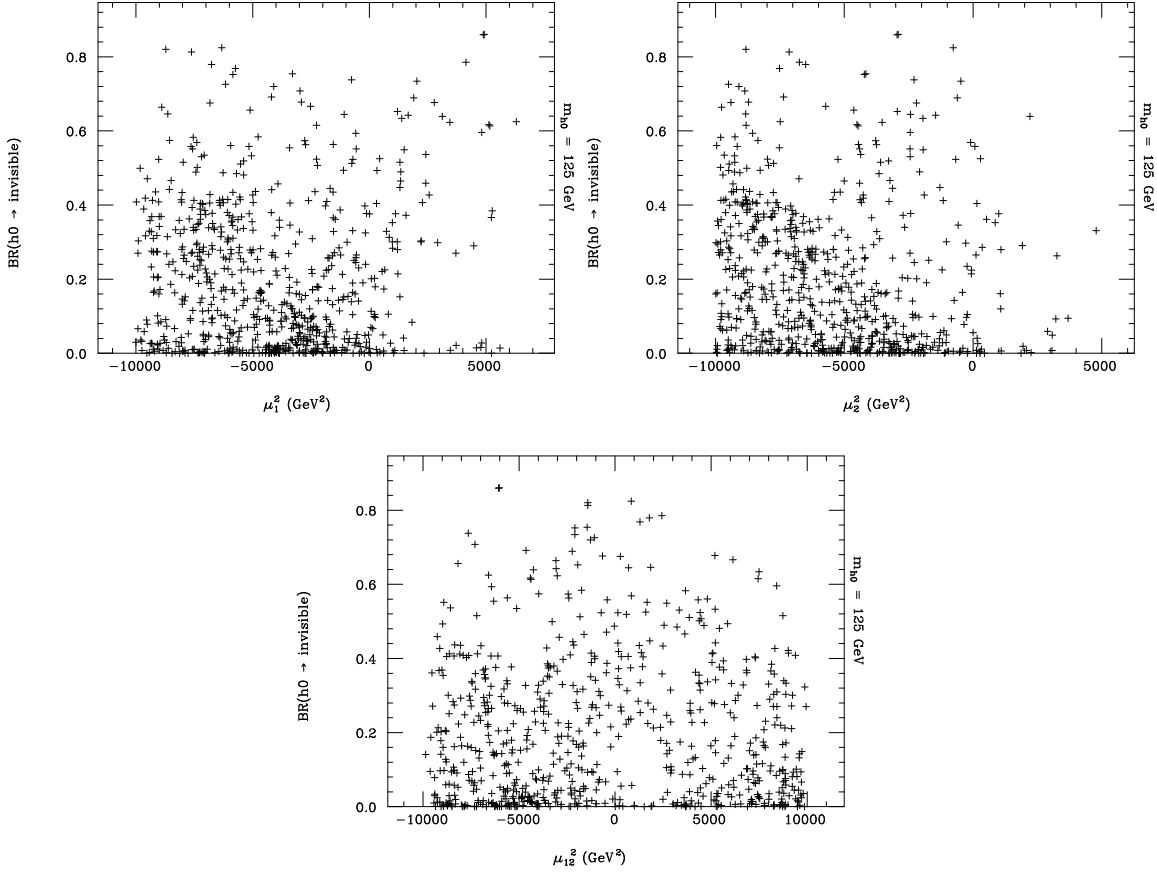

\centering
\includegraphics[width=0.35\linewidth,angle=90]{./3HDM/mu1_2-BR-invisible.ps}
\includegraphics[width=0.35\linewidth,angle=90]{./3HDM/mu2_2-BR-invisible.ps}\\[0.5cm]
\includegraphics[width=0.35\linewidth,angle=90]{./3HDM/mu12_2-BR-invisible.ps}
\caption{Parameter scan of the I(2+1)HDM mapped in terms of 
BR$(h\to{H_{1}}{H_{1}})$ as a function of
${\mu}_1^2$ (top-left),
${\mu}_2^2$ (top-right) and
${\mu}_{12}^2$ (bottom).}
\label{fig:3HDM-params1}
\end{figure}

\begin{figure}[!t]
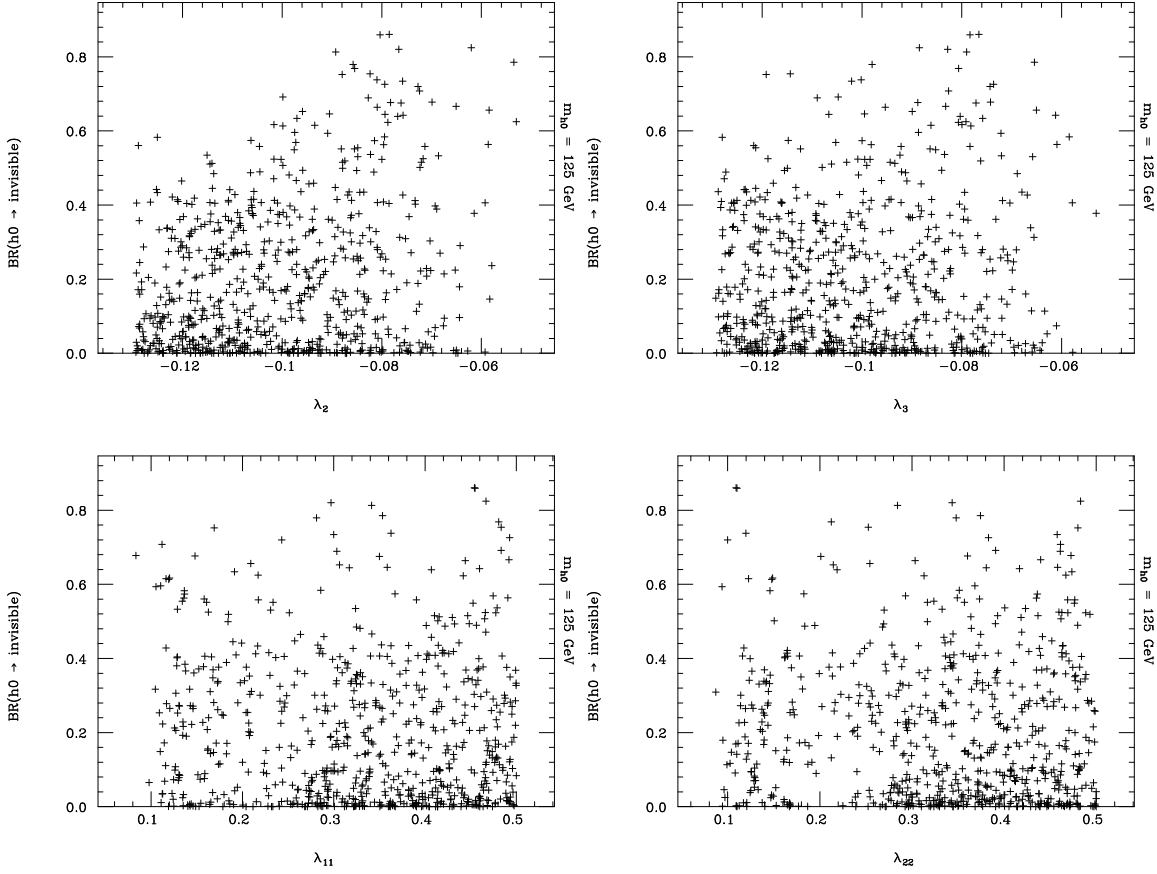

\centering
\includegraphics[width=0.35\linewidth,angle=90]{./3HDM/lam2-BR-invisible.ps}
\includegraphics[width=0.35\linewidth,angle=90]{./3HDM/lam3-BR-invisible.ps}\\[0.5cm]
\includegraphics[width=0.35\linewidth,angle=90]{./3HDM/lam11-BR-invisible.ps}
\includegraphics[width=0.35\linewidth,angle=90]{./3HDM/lam22-BR-invisible.ps}
\caption{Parameter scan of the I(2+1)HDM mapped in terms of 
BR$(h\to{H_{1}}{H_{1}})$ as a function of
${\lambda}_2$ (top-left),
${\lambda}_3$ (top-right),
${\lambda}_{11}$ (bottom-left) and
${\lambda}_{22}$ (bottom-right).}
\label{fig:3HDM-params2}
\end{figure}

\begin{figure}[!h]
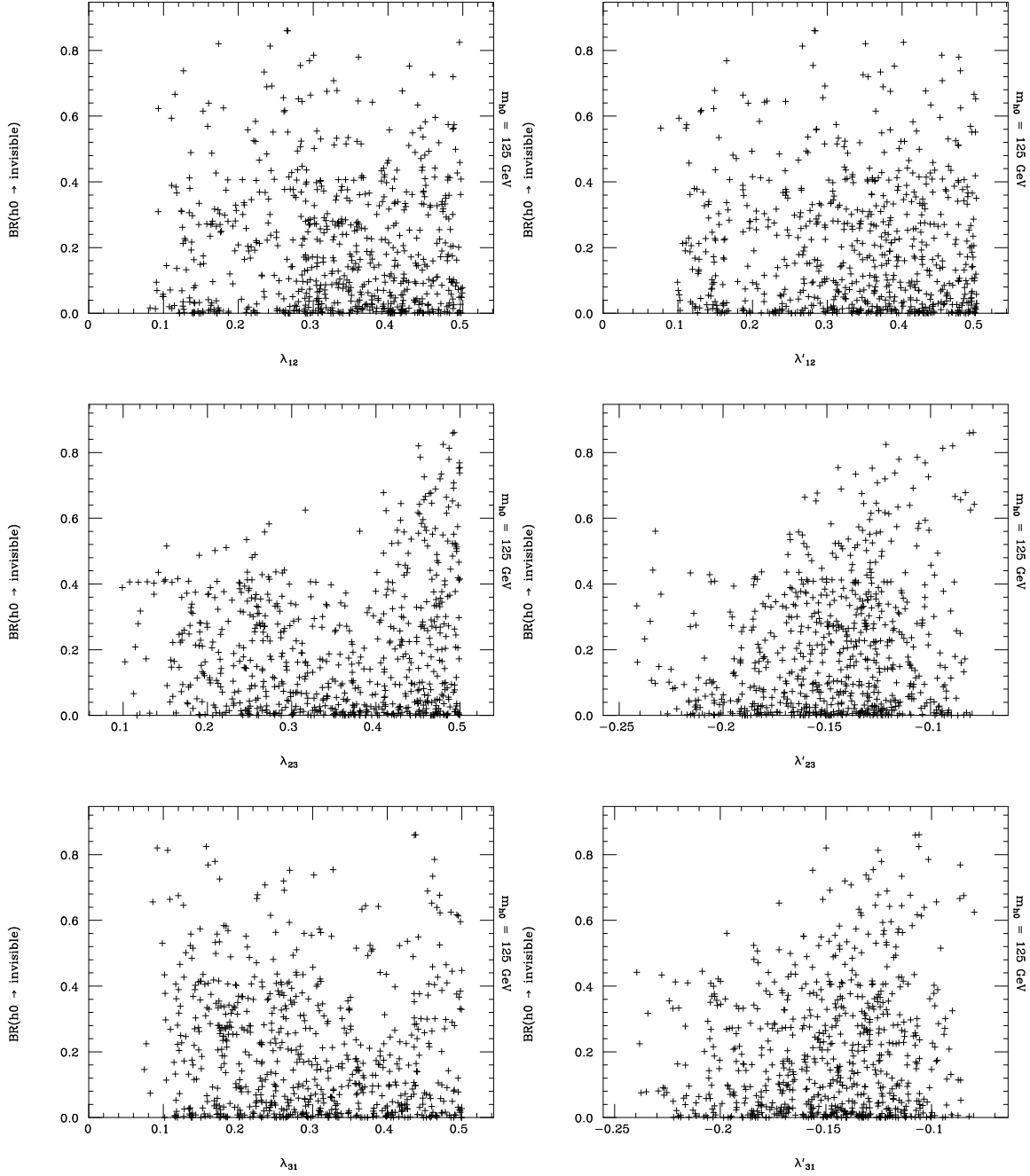

\centering
\includegraphics[width=0.35\linewidth,angle=90]{./3HDM/lam12-BR-invisible.ps}
\includegraphics[width=0.35\linewidth,angle=90]{./3HDM/lan12-BR-invisible.ps}\\[0.5cm]
\includegraphics[width=0.35\linewidth,angle=90]{./3HDM/lam23-BR-invisible.ps}
\includegraphics[width=0.35\linewidth,angle=90]{./3HDM/lan23-BR-invisible.ps}\\[0.5cm]
\includegraphics[width=0.35\linewidth,angle=90]{./3HDM/lam31-BR-invisible.ps}
\includegraphics[width=0.35\linewidth,angle=90]{./3HDM/lan31-BR-invisible.ps}
\caption{Parameter scan of the I(2+1)HDM mapped in terms of 
BR$(h\to{H_{1}}{H_{1}})$ as a function of
${\lambda }_{12}$ (top-left),
${\lambda'}_{12}$ (top-right),
${\lambda }_{23}$ (middle-left),
${\lambda'}_{23}$ (middle-right),
${\lambda }_{31}$ (bottom-left) and
${\lambda'}_{31}$ (bottom-right).}
\label{fig:3HDM-params3}
\end{figure}

This is done in Figs.~(\ref{fig:3HDM-params1}--\ref{fig:3HDM-params3}). The distributions of the points surviving the theoretical constraints described in the previous subsection and the experimental 
ones in Eq.~(\ref{eq:3HDM-exp_constraints}), which limit the physically acceptable values of the input parameters of the model as seen along the 
$x$-axis in the plots, is rather uniform, reaching up to a value of the invisible Higgs BR of nearly 
90\%, with the majority of data accumulating below the 20\% mark. Hence, in the light of current experimental constraints
on this variable, which place an upper value for it at approximately
$40\%$ from direct searches \cite{invisible-direct} and  20\% or so from indirect ones  \cite{invisible-indirect,Cheung:2014noa}, we would
conclude that the I(2+1)HDM realisation discussed here is still compatible with the corresponding LHC data. Furthermore, owing to the  (somewhat) increased density of points towards the lower end of the BR distribution, one should expect upcoming data from the second run of the
CERN collider to closely scrutinise this model.

\begin{figure}[!h]
\centering
\includegraphics[width=0.75\linewidth,angle=0]{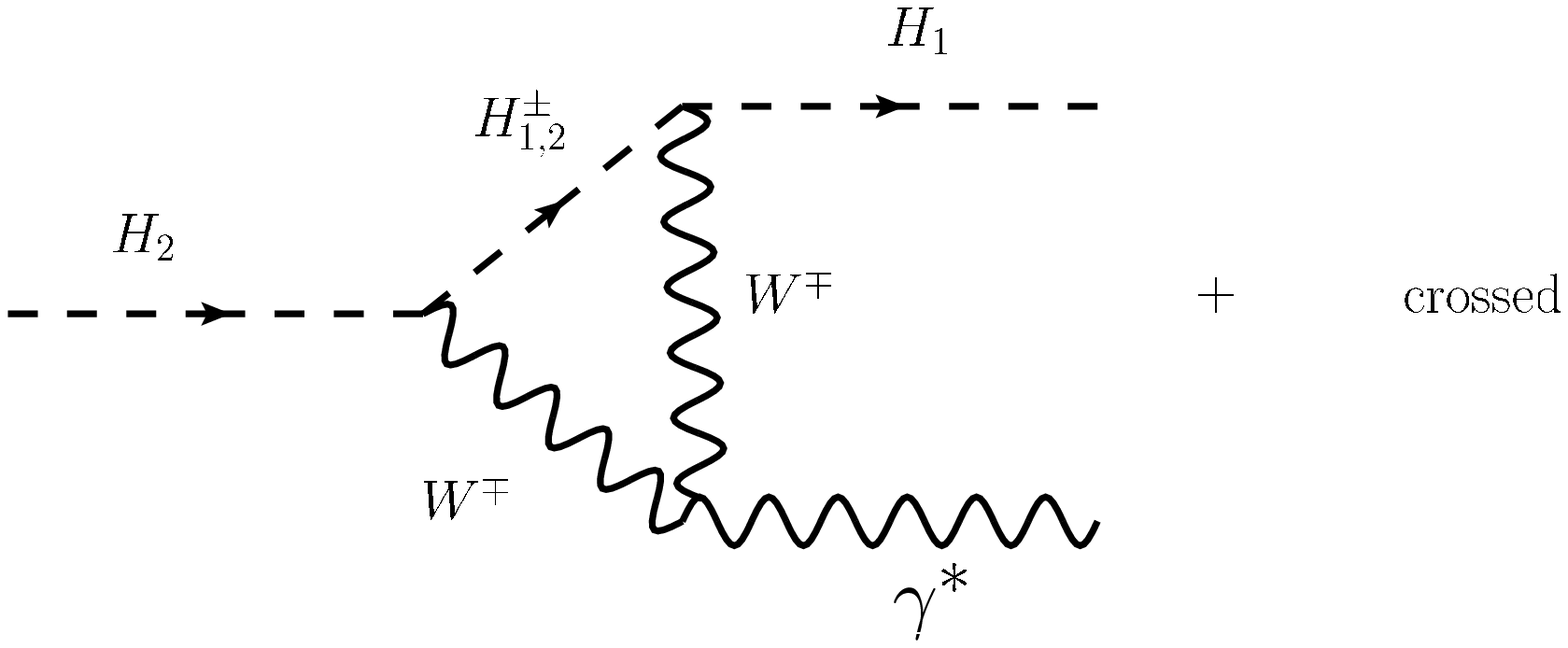}
\vspace*{1.0cm}
\caption{The loop diagrams responsible for $H_2\rightarrow H_1\gamma^*$ decays,
where $H_1$ is absolutely stable and hence invisible,
while $\gamma^*$ is a virtual photon that couples to fermion-antifermion pairs.}
\label{loop}
\end{figure}

\begin{figure}[!t]
\centering
\includegraphics[width=0.50\linewidth,angle=90]{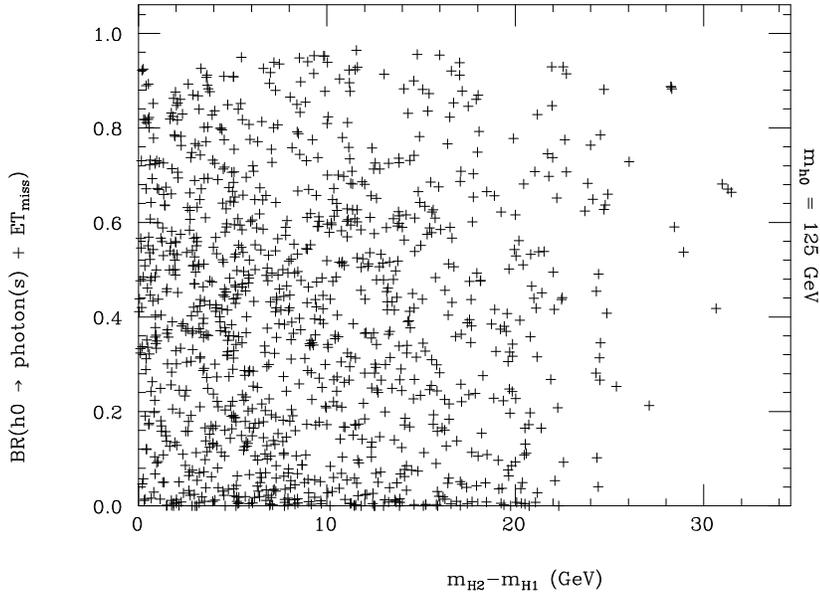}
\caption{Parameter scan of the I(2+1)HDM mapped in terms of 
BR$(h\to{H_{1}}{H_{2}})$ + BR$(h\to{H_{2}}{H_{2}})$ as a function of
$m_{H_2}-m_{H_1}$. The plot indicates SM Higgs decay rates into photon(s) of up to about 30 GeV 
plus missing (transverse) energy.}
\label{fig:3HDM-masses-radiative}
\end{figure}

However, we believe that the most striking signals for our I(2+1)HDM would actually be the ones induced by radiative decays
of heavy inert Higgs states into the DM candidate. It turns out from our scan that the primary $h$ decay channels inducing
such a pattern are $h\to  {H_{1}}{H_{2}}$ and $h\to  {H_{2}}{H_{2}}$, wherein the ${H_{2}}$ state decays into ${H_{1}}\gamma$ (with the photon being off-shell, thus yielding $e^+ e^-$ pairs) via a triangle-loop
involving off-shell $H^\pm_{1,2} W^\mp$ pairs with 100\% probability
(see Fig.~(\ref{loop})).

In fact, the mass spectrum emerging in the
inert Higgs sector prevents other possible chained decays involving the other inert Higgs states (i.e.,  ${A_{1}}$,
${H^\pm_{1}}$, ${A_{2}}$ and ${H^\pm_{2}}$) as the rest masses of the latter are such that none of the other $h$-inert-Higgs couplings listed in Appendix A can onset any other $h$ primary decay into combinations of pairs of these heavier inert states. (In fact, the $H_1$ and  $H_2$ states are always the lightest  ones amongst the inert doublet ones.)  
Fig.~(\ref{fig:3HDM-masses-radiative}) shows the cumulative BR of the entire decay chains 
$h\to  {H_{1}}{H_{2}}\to H_1H_1\gamma$ and $h\to  {H_{2}}{H_{2}}\to H_1H_1\gamma\gamma$ as a function of the ${H_2}$ and ${H_1}$ mass difference. 
In the end, such decay chains could result in highly energetic Electro-Magnetic (EM) showers, one or two at a time (alongside significant missing (transverse) energy, $E^T_{\rm miss}$, induced by the DM pair),
which would generally be captured by the
detectors, as the mass difference $m_{H_2}-m_{H_1}$ can reach up to 30 GeV or so.
The new SM Higgs decay channels into off-shell photon(s) plus missing energy, enables the I(2+1)HDM 
to be distinguished from the I(1+1)HDM, as CP-conservation prevents such radiative decays in its inert sector.

\section{The I(4+2)HDM with $ \tan \beta = 1$}
\label{6HDMtb1}

\subsection{A highly symmetric potential}
In a 6HDM, the most general phase invariant part of the potential has the following form
\be
V_0 = \sum^6_i \left[ \mu^2_i (\phi_i^\dagger \phi_i) + \lambda_{ii} (\phi_i^\dagger \phi_i)^2\right] 
+ \sum^6_{ij}\left[\lambda_{ij}(\phi_i^\dagger \phi_i) (\phi_j^\dagger \phi_j) + 
\lambda'_{ij}(\phi_i^\dagger \phi_j) (\phi_j^\dagger \phi_i)\right]. 
\ee
To reduce the number of free parameters in this potential, we make use of the resemblance of this model to the I(1+1)HDM and imitate its potential structure.

The general I(1+1)HDM potential has the following form
\bea
\label{IDM-pot}
V^{I(1+1)HDM} &=& - \mu^2_{i} (\phi_i^\dagger \phi_i) -\mu^2_a (\phi_a^\dagger \phi_a)  \\
&&+ \lambda_{ii} (\phi_i^\dagger \phi_i)^2+ \lambda_{aa} (\phi_a^\dagger \phi_a)^2  \nonumber\\
&&+ \lambda_{ai}  (\phi_a^\dagger \phi_a)(\phi_i^\dagger \phi_i) + \lambda'_{ai} (\phi_a^\dagger \phi_i)(\phi_i^\dagger \phi_a)  \nonumber\\ 
&&+ \tilde{\lambda}(\phi_i^\dagger\phi_a)^2  + h.c. \nonumber
\eea
where $a$ and $i$ identify \textit{active} and \textit{inert} doublets respectively.

We construct our potential by replacing the active doublet, $\phi_a$, with our two active doublets $\phi_5,\phi_6$, and the inert doublet, $\phi_i$, with our four inert doublets $\phi_1,\phi_2,\phi_3,\phi_4$. The phase invariant part of the potential, $V_0$, gets the form:
\bea
\label{V0-part-1}
V_0 &=& \mu^2_a \biggl[ \phi_5^\dagger \phi_5 + \phi_6^\dagger \phi_6 \biggr] + \mu^2_i \biggl[\phi_1^\dagger \phi_1 + \phi_2^\dagger \phi_2 + \phi_3^\dagger \phi_3 + \phi_4^\dagger \phi_4 \biggr] \\
&&+ \lambda_{a} \biggl[\phi_5^\dagger \phi_5 + \phi_6^\dagger \phi_6 \biggr]^2 + \lambda_{i} \biggl[\phi_1^\dagger \phi_1 + \phi_2^\dagger \phi_2 + \phi_3^\dagger \phi_3 + \phi_4^\dagger \phi_4 \biggr]^2  \nonumber\\
&&+ \lambda_{ai} \biggl[(\phi_5^\dagger \phi_5 + \phi_6^\dagger \phi_6)(\phi_1^\dagger \phi_1 + \phi_2^\dagger \phi_2 + \phi_3^\dagger \phi_3 + \phi_4^\dagger \phi_4) \biggr]  \nonumber\\
&&+ \lambda'_{ai} \biggl[(\phi_5^\dagger \phi_1)(\phi_1^\dagger \phi_5) + (\phi_5^\dagger \phi_2)(\phi_2^\dagger \phi_5) + (\phi_5^\dagger \phi_3)(\phi_3^\dagger \phi_5) + (\phi_5^\dagger \phi_4)(\phi_4^\dagger \phi_5) \nonumber\\
&& \qquad \quad + (\phi_6^\dagger \phi_1)(\phi_1^\dagger \phi_6) + (\phi_6^\dagger \phi_2)(\phi_2^\dagger \phi_6) + (\phi_6^\dagger \phi_3)(\phi_3^\dagger \phi_6) + (\phi_6^\dagger \phi_4)(\phi_4^\dagger \phi_6) \biggr]. \nonumber
\eea

{Note the absence of a quartic term mixing the active doublets which will lead to a mass degeneracy between the scalars resulting from the active sector, namely $H$, $A$ and $H^\pm$ fields (see the mass spectra in Eq.~(\ref{mass-simple}) and Eq.~(\ref{mass-general})). To remove this degeneracy the following phase invariant term could be added to the potential:
\be 
\label{V0-part-2}
\lambda_{aa} (\phi^\dagger_5 \phi_6 + \phi^\dagger_5 \phi_6 )^2.
\ee
}
Constructing the $Z_2$-symmetric part of the potential depends on the generator of the group. Inspired by E$_6$SSM, we impose a $Z^H_2$ symmetry generated by:
\be 
g_{Z^H_{2}}=  \mathrm{diag}(-1, -1, -1, -1, +1, +1 ).
\ee
The terms that are needed to be added to $V_0$ in Eq.~(\ref{V0-part-1})--(\ref{V0-part-2}) which ensure the $Z_2$ symmetry of the potential are the following:
\be 
V_{Z_2} = \lambda_{1} \biggl[(\phi^\dagger_1 + \phi^\dagger_2 + \phi^\dagger_3 + \phi^\dagger_4)( \phi_5 + \phi_6)\biggr]^2 +h.c. \label{Z2-part}
\ee
Note that the VEV alignment $(0,0,0,0,\frac{v}{\sqrt{2}},\frac{v}{\sqrt{2}})$ respects this symmetry.

Constructing the potential in this way leads to extra symmetries of the potential. The phase invariant part has an $SO(4) \times SO(6)$ symmetry. The $V_{Z_2}$ part is also symmetric under the exchange of $\phi_5$ and $\phi_6$ as well as under any permutation of $\phi_1,\phi_2,\phi_3,\phi_4$. Therefore, the true symmetry of the full potential is $Z^H_2 \times Z^{\phi_5 \leftrightarrow \phi_6}_2 \times S_4$.

These extra symmetries could be removed by adding the following terms with complex coefficients:
\bea 
&& {\mu'_a}^2(\phi^\dagger_5 \phi_6) +{\mu'_i}^2 \left[\phi^\dagger_1 (\phi_2 + \phi_3 + \phi_4) + \phi^\dagger_2 (\phi_3 + \phi_4) + \phi^\dagger_3 \phi_4 \right] + h.c. 
\label{Z2-S4-removing}
\eea
These terms also introduce mixing between the active doublets and between the inert doublets which would prevent the appearance of extra massless scalars.

Note that ${\mu'_a}^2$ and ${\mu'_i}^2$ have to be complex in order to break the $Z^{\phi_5 \leftrightarrow \phi_6}_2$ and $S_4$ symmetries. However, as it will be shown in section \ref{minimization}, the VEV alignment $(0,0,0,0,v,v)$ is not a minimum of the potential if ${\mu'_a}^2$ is complex. Thus we require ${\mu'_a}^2$ to be real and therefore the $Z^{\phi_5 \leftrightarrow \phi_6}_2$ symmetry remains unbroken.

Allowing for complex ${\mu'_i}^2$ (and other parameters, namely $\lambda_1$ and $\lambda'_1$) leads to explicit CP-violation in the model. We shall not consider CP-violation in this paper. Therefore, we require all parameters to be real, leading to the original $Z^H_2 \times Z^{\phi_5 \leftrightarrow \phi_6}_2 \times S_4$ symmetry of the potential.

\subsection{A DM candidate}
In this model the DM candidate is the lightest neutral $Z^H_2$ odd particle among others resulting from the four inert doublets. However, we shall impose an extra $Z^{DM}_2$ with a generator of the form
\be 
g_{Z^{DM}_2}=  \mathrm{diag}(-1, +1, +1, +1, +1, +1 )
\ee 
to protect the DM candidate which is the neutral state from $\phi_1$, 
%(which is $H^0_{1u}$ in the notation we introduce in section \ref{notation}), 
from decaying into other $Z^H_2$ odd particles which are the fields from $\phi_2$, $\phi_3$ and $\phi_4$ doublets which are now protected by an $S_3$ symmetry amongst themselves. 
Imposing the $Z^{DM}_2$ symmetry reduces Eq.~(\ref{Z2-part}) and Eq.~(\ref{Z2-S4-removing}) to
\bea 
\label{Z2-DM}
&& {\mu'_a}^2(\phi^\dagger_5 \phi_6) + {\mu'_i}^2 \left[ \phi^\dagger_2 (\phi_3 + \phi_4) + \phi^\dagger_3 \phi_4 \right] + \\
&& \lambda_{1} \biggl[(\phi^\dagger_2 + \phi^\dagger_3 + \phi^\dagger_4)( \phi_5 + \phi_6)\biggr]^2 + \lambda'_1 \left[\phi^\dagger_1 (\phi_5 + \phi_6) \right]^2 +   h.c. \nonumber
\eea
The resulting potential consistent of $V_0$ in Eq.~(\ref{V0-part-1}), Eq.~(\ref{V0-part-2}) and Eq.~(\ref{Z2-DM}) is symmetric under the $Z^H_2 \times Z^{\phi_5 \leftrightarrow \phi_6}_2 \times S_3 \times Z^{DM}_2$ group.

Note that the fields from the three inert doublets, $\phi_2$, $\phi_3$ and $\phi_4$, are also protected by their unbroken $S_3$ symmetry and the lightest Higgs state in this sector will be absolutely stable,
providing a separate candidate for DM, in addition to the lightest Higgs state from $\phi_1$.
The motivation for introducing the $Z^{DM}_2$ is to preserve a DM candidate from $\phi_1$,
even if the $Z^H_2$ is broken. Note that, in the E$_6$SSM, $Z^H_2$ is not an exact symmetry
\cite{King:2005jy,King:2005my}. If $Z^{DM}_2$ were not imposed then all the four inert doublets could mix,
leading to a larger variety of the possible cascade decays that we will shortly discuss.

%%%%%%%%%%%%%%%%%%%%%%%%%%%%%%%%%%%%%%%%%%%%%%%%%%%%%%%%%%%%%%%%%%%%%%%%%%%%%%%%%%%%%%%%%%%%%
%\subsection{The case $\tan\beta =1$}\label{minimization}

\subsection{The highly symmetric potential in E$_6$SSM notation}\label{notation}
To write the potential in the language of the E$_6$SSM, we define
\bea 
H_{1u} \equiv \phi_1 , \quad   H_{2u}\equiv \phi_3 , \quad H_{3u}\equiv \phi_5   \\
H_{1d} \equiv \phi_2 , \quad   H_{2d}\equiv \phi_4 , \quad H_{3d}\equiv \phi_6  \nonumber
\label{ESSM-language}
\eea
where
\be 
H_{\alpha u}: \doublet{$\begin{scriptsize}$H^+_{\alpha u}$\end{scriptsize}$}{\frac{H^0_{\alpha u}+iA^0_{\alpha u}}{\sqrt{2}}}, \qquad 
H_{\alpha d}: \doublet{$\begin{scriptsize}$H^+_{\alpha d}$\end{scriptsize}$}{\frac{H^0_{\alpha d}+iA^0_{\alpha d}}{\sqrt{2}}}  \nonumber
\ee
and $\alpha=1,2,3$.

The full  $Z^H_2 \times Z^{\phi_5 \leftrightarrow \phi_6}_2 \times S_3 \times Z^{DM}_2$ potential written in terms of the new fields has the following form:
\bea
\label{simple-potential}
V &=& \mu^2_a \biggl( |H_{3u}|^2 + |H_{3d}|^2 \biggr) + \mu^2_i \biggl(|H_{1u}|^2 + |H_{1d}|^2 + |H_{2u}|^2 + |H_{2d}|^2 \biggr) \\
&&+ \mu'^2_a \biggl(H^\dagger_{3u} H_{3d}  \biggr) + \mu'^2_i \biggl(H^\dagger_{1d} H_{2u} + H^\dagger_{1d} H_{2d} + H^\dagger_{2u} H_{2d}   \biggr) + h.c. \nonumber\\
&&+ \lambda_a \biggl( |H_{3u}|^2 + |H_{3d}|^2 \biggr)^2 +\lambda_i \biggl( |H_{1u}|^2 + |H_{1d}|^2 + |H_{2u}|^2 + |H_{2d}|^2 \biggr)^2 \nonumber \\
&&+ \lambda_{ai} \biggl[ ( |H_{3u}|^2 + |H_{3d}|^2)(|H_{1u}|^2 + |H_{1d}|^2 + |H_{2u}|^2 + |H_{2d}|^2)   \biggr] \nonumber\\
&&+ \lambda'_{ai} \biggl[|H^\dagger_{3u}H_{1u}|^2+  |H^\dagger_{3u}H_{1d}|^2+ |H^\dagger_{3u}H_{2u}|^2+ |H^\dagger_{3u}H_{2d}|^2 \nonumber\\
&& \qquad \quad + |H^\dagger_{3d}H_{1u}|^2+  |H^\dagger_{3d}H_{1d}|^2+ |H^\dagger_{3d}H_{2u}|^2+ |H^\dagger_{3d}H_{2d}|^2    \biggr] \nonumber\\
&&  {+ \lambda_{aa} \biggl[H^\dagger_{3u}H_{3d} + H^\dagger_{3d}H_{3u} \biggr]^2 } \nonumber\\
&&+ \lambda_{1} \biggl[ H^\dagger_{1d}  H_{3u} + H^\dagger_{1d} H_{3d} + H^\dagger_{2u}H_{3u} + H^\dagger_{2u} H_{3d}+ H^\dagger_{2d}H_{3u} + H^\dagger_{2d}H_{3d} \biggr]^2 +h.c. \nonumber\\
&&+ \lambda'_{1} \biggl[  H^\dagger_{1u} H_{3u}  + H^\dagger_{1u} H_{3d} \biggr]^2 + h.c. \nonumber
\eea

\subsection{The mass spectrum}\label{minimization}
For this highly symmetric potential the vacuum alignment is of the form
\be 
\langle H_{1u} \rangle =\langle H_{1d} \rangle =\langle H_{2u}\rangle =\langle H_{2d} \rangle =0, \qquad \langle H_{3u}\rangle =\langle H_{3d}\rangle =\frac{v}{\sqrt{2}}.
\ee
Expanding the potential around this point leads to the following linear terms:
\bea
\label{linear-terms}
V^{(1)} &=& \left[ v_u \mu^2_a + v_d \Re{\mu'^2_a} +  v_u(v^2_u+ v^2_d)\lambda_a + {2v_u v^2_d \lambda_{aa}} \right] H^0_{3u} \\
&+&  \left[ v_d \mu^2_a + v_u \Re{\mu'^2_a} +  v_d(v^2_u+ v^2_d)\lambda_a + {2v_d v^2_u \lambda_{aa}} \right] H^0_{3d} \nonumber\\
&-& i\left[ v_d \Im{\mu'^2_a} \right] A^0_{3u} + i \left[ v_u \Im{\mu'^2_a} \right] A^0_{3d} . \nonumber
\eea
For this point to be a minimum of the potential the linear terms must vanish, therefore
\bea 
&& \bullet \quad \Im{\mu'^2_a} = 0 \\
&& \bullet \quad  v_u \left(\mu^2_a + \lambda_a (v^2_u+ v^2_d) +{2v^2_d \lambda_{aa}} \right) + v_d \Re{\mu'^2_a} = 0 \nonumber\\[2mm]
&& \bullet \quad v_d \left(\mu^2_a + \lambda_a (v^2_u+ v^2_d) +{2v^2_u \lambda_{aa}} \right) + v_u \Re{\mu'^2_a} = 0 \nonumber
\eea
Satisfying these equations simultaneously requires $\mu'^2_a$ to be real and results in $v^2_u = v^2_d$ or equivalently $v_u = \pm v_d$ leading to $\tan\beta = \pm 1$, where $\tan \beta = v_u/v_d$. In this section we study the $\tan\beta = 1$ case.

By construction, the above vacuum alignment respects the symmetry of the potential. 
 Expanding the potential around this point results in 
\be 
v^2 = \frac{-\mu^2_a - \mu'^2_a}{ 2\lambda_a {+2\lambda_{aa}}} 
\ee
and the following mass spectrum:

\begin{footnotesize}
\bea
\label{mass-simple}
&& \textbf{h} =\frac{H^0_{3u}+H^0_{3d}}{\sqrt{2}}: 
\quad m^2=-2\mu^2_a-2\mu'^2_a  \\
&& \textbf{H}=\frac{H^0_{3u}- H^0_{3d}}{\sqrt{2}}:
\quad m^2=-2\mu'^2_a  \nonumber\\
&& \textbf{G$^\pm$} =\frac{H^\pm_{3u}+H^\pm_{3d}}{\sqrt{2}}:
\quad m^2=0 \nonumber\\
&& \textbf{H$^\pm$}=\frac{H^\pm_{3u}-H^\pm_{3d}}{\sqrt{2}}: 
\quad m^2=-2\mu'^2_a {-4\lambda_{aa}v^2} \nonumber\\
&& \textbf{G$^0$} =\frac{A^0_{3u}+A^0_{3d}}{\sqrt{2}}:
\quad m^2=0 \nonumber\\
&& \textbf{A}=\frac{A^0_{3u}-A^0_{3d}}{\sqrt{2}}:
\quad m^2=-2\mu'^2_a {-4\lambda_{aa}v^2} \nonumber\\
&& \textbf{H$^{\pm}_{1u}$}  : 
\quad m^2={\mu_i}^2+\lambda_{ai} v^2 \nonumber\\[2mm]
&& \textbf{H$^0_{1u}$} : 
\quad m^2=\mu^2_i + (\lambda_{ai}+ \lambda'_{ai} + 4\lambda'_1) v^2 \nonumber\\[2mm]
&& \textbf{A$^0_{1u}$} : 
\quad m^2=\mu^2_i + (\lambda_{ai}+ \lambda'_{ai} - 4\lambda'_1) v^2 \nonumber\\[2mm]
&& \textbf{H$_x$} = \frac{H^0_{1d}-H^0_{2d}}{\sqrt{2}}: 
\quad m^2=\mu^2_i- \mu'^2_i + (\lambda_{ai}+ \lambda'_{ai})v^2 \nonumber\\[2mm]
&& \textbf{H$_y$}=\frac{H^0_{1d}-H^0_{2u}}{\sqrt{2}}: 
\quad m^2=\mu^2_i- \mu'^2_i + (\lambda_{ai}+ \lambda'_{ai})v^2 \nonumber\\[2mm]
&& \textbf{H$_z$} =\frac{H^0_{1d}+H^0_{2u}+H^0_{2d}}{\sqrt{3}}:  
\quad m^2=\mu^2_i+ 2\mu'^2_i + (\lambda_{ai}+ \lambda'_{ai}+12\lambda_1)v^2 \nonumber\\
&& \textbf{A$_x$}=\frac{A^0_{1d}-A^0_{2d}}{\sqrt{2}}: 
\quad m^2=\mu^2_i- \mu'^2_i + (\lambda_{ai}+ \lambda'_{ai})v^2 \nonumber\\[2mm]
&& \textbf{A$_y$}=\frac{A^0_{1d}-A^0_{2u}}{\sqrt{2}}: 
\quad m^2=\mu^2_i- \mu'^2_i + (\lambda_{ai}+ \lambda'_{ai})v^2 \nonumber\\[2mm]
&& \textbf{A$_z$}=\frac{A^0_{1d}+A^0_{2u}+A^0_{2d}}{\sqrt{3}}: 
\quad m^2= \mu^2_i+ 2\mu'^2_i + (\lambda_{ai}+ \lambda'_{ai} -12\lambda_1)v^2 \nonumber\\
&& \textbf{H$^\pm_x$} = \frac{H^{\pm}_{1d}-H^{\pm}_{2d}}{\sqrt{2}}: 
\quad m^2=\mu^2_i- \mu'^2_i + \lambda_{ai}v^2 \nonumber\\[2mm]
&& \textbf{H$^\pm_y$} =\frac{H^{\pm}_{1d}-H^{\pm}_{2u}}{\sqrt{2}}: 
\quad m^2= \mu^2_i- \mu'^2_i + \lambda_{ai}v^2 \nonumber\\[2mm]
&& \textbf{H$^\pm_z$} =\frac{H^{\pm}_{1d}+H^{\pm}_{2u}+H^{\pm}_{2d}}{\sqrt{3}}: 
\quad  m^2=\mu^2_i+ 2\mu'^2_i + \lambda_{ai}v^2 \nonumber
\eea
\end{footnotesize}

%\begin{center}
%\includegraphics[scale=0.9]{Masses-6HDM-tanb1.ps} 

%\end{center}

Note that the base fields from the first doublet ($H^0_{1u},A^0_{1u},H^\pm_{1u}$) are mass eigenstates and the neutral fields $H^0_{1u}$ and $A^0_{1u}$ are the DM candidates in this model.

The Feynman rules for this potential are presented in Appendix \ref{simple}. 
%The results can be used for the $v_u = - v_d$ case when $\tan\beta = 1$ is replaced by $\tan\beta = - 1$. 

\subsection{Constraints on the parameters}
As it will be shown in the Feynman rules, the state $h$ is the only field that couples to $W^\pm$ and $Z$ gauge bosons and therefore plays the role of the SM-Higgs. 
Furthermore, imposing $m^2_{h} < m^2_{H}$ leads to
\be 
\bullet \quad \mu^2_a>0.  
\ee
We also require the neutral DM candidates decay channel to the charged field from the $H_{1u}$ doublet to be closed, therefore:
\bea 
&&\bullet~ \mbox{if $H^0_{1u}$ is DM the candidate:} \quad m^2_{H^0_{1u}} < 2 m^2_{H^\pm_{1u}} \quad \rightarrow \quad   (4\lambda'_1 +  \lambda'_{ai})v^2  < \mu^2_i + \lambda_{ai} v^2 \nonumber\\
&&\bullet~ \mbox{if $A^0_{1u}$ is the DM candidate:} \quad m^2_{A^0_{1u}} < 2 m^2_{H^\pm_{1u}} \quad \rightarrow \quad  (-4\lambda'_1 +  \lambda'_{ai})v^2  <\mu^2_i + \lambda_{ai} v^2 \nonumber
\eea

Further constraints on the parameters are required by:
\begin{enumerate}
\item 
\textbf{Positivity of the mass eigenstates}:
\bea
&& \bullet \quad \lambda_a + \lambda_{aa} > 0  \\
&& \bullet \quad \mu^2_a + 2\lambda_{a} v^2 > 0 \nonumber\\
&& \bullet \quad \mu^2_a + 2(\lambda_a + \lambda_{aa}) v^2 > 0 \nonumber\\
&& \bullet \quad \mu^2_i + \lambda_{ai} v^2 > 0 \nonumber\\
&& \bullet \quad \mu^2_i + (\lambda_{ai}+ \lambda'_{ai}) v^2  > \mu'^2_i \nonumber\\
&& \bullet \quad \mu^2_i + (\lambda_{ai}+ \lambda'_{ai}) v^2  >  4 |\lambda'_1 |v^2 \nonumber\\
&& \bullet \quad \mu^2_i +2\mu'^2_i + (\lambda_{ai}+ \lambda'_{ai}) v^2 >  12 |\lambda_1| v^2  \nonumber
\eea

\item
\textbf{Bounded-ness of the potential}:
\bea
&& \bullet \quad  \lambda_{a} > 0 \\
&& \bullet \quad  \lambda_i   >0    \nonumber\\
&& \bullet \quad  \lambda_{ai} + \lambda'_{ai} > -2 \sqrt{\lambda_{a}\lambda_{i}} \nonumber\\
&& {\bullet \quad 4\lambda_a + \lambda_{aa} > 0} \nonumber
\eea
and we require the parameters of the $V_{Z_2}$ part to be smaller than the parameters of the $V_0$ part of the potential:
\be 
\bullet \quad |\lambda_1|, |\lambda'_1| < |\lambda_a|, |\lambda_i|, |\lambda_{ai}|, |\lambda'_{ai}|, {|\lambda_{aa}|}
\ee

\item
\textbf{Positive-definite-ness of the Hessian}:
\bea
&& \bullet \quad \mu^2_i + (\lambda_{ai}+ \lambda'_{ai}) v^2  > 0 \\
&& \bullet \quad \mu^2_i + (\lambda_{ai}+ \lambda'_{ai}) v^2  > \mu'^2_i \nonumber\\
&& \bullet \quad \mu^2_i + (\lambda_{ai}+ \lambda'_{ai}) v^2  > -2\mu'^2_i \nonumber\\
&& \bullet \quad (-\mu'^2_a+\lambda_a v^2 )^2  > (\mu'^2_a + (\lambda_a+6\lambda_{aa}) v^2)^2  \nonumber
\eea 

\end{enumerate}

\subsection{Results for the I(4+2)HDM with $\tan \beta = 1$ }

The case $\tan\beta=1$ implies that the nature of the lightest Higgs state in the spectrum, $h$, is such that its coupling
to all SM matter are identical to those of the SM Higgs boson, similar to what we have already seen for the I(2+1)HDM
case. Hence, again, the only effects which could ensue from the enlarged
Higgs sector must be found in the interactions between the $h$ state and its new Higgs partners. Specifically, 
the hallmark signal would be an increased invisible width of the 
SM-like $h$ state, owing to the possibility of the following decays: either
$h\to {H^0_{1u}}{H^0_{1u}}$ or $h\to {A^0_{1u}}{A^0_{1u}}$, wherein the ${H^0_{1u}}$ or ${A^0_{1u}}$
is the DM candidate, respectively, which is stable and escapes the detectors. In fact,  when one of two channels is open the other is not, owing to the
fact if $m_h>2 m_{H^0_{1u}}$ then $m_h<2 m_{A^0_{1u}}$ and vice versa\footnote{In essence, the splitting between the squared masses of $H^0_{1u}$ and $A^0_{1u}$ is $\pm 8\lambda'_1 v^2$ and, depending on the sign of $\lambda'_1$, either $H^0_{1u}$ or $A^0_{1u}$ is the DM candidate.}.
In order to quantify the latter we have performed a scan over the 9-dimensional parameter space of the model, by randomly 
generating points over the following intervals\footnote{Notice that we have fixed
$\lambda_a$ in terms of $m_h=125$ GeV.}:
$$
        0 < {\mu}_a^2 < 10~{\rm TeV}^2,  
$$
$$
        -10~{\rm TeV}^2 < {\mu'}_a^2 < 0,  
$$
$$
        -10~{\rm TeV}^2 < {\mu }_i^2 , {\mu'}_i^2 < 10~{\rm TeV}^2, 
$$
$$
      0     <   \lambda_i              < 0.5,
$$
$$
     -0.5  <  \lambda_{ai}   <0.5,
$$
\begin{equation}\label{eq:the_constraints}
     -0.5  <  \lambda'_{ai}  ,  \lambda_{1} ,   \lambda'_{1}  < 0.5,
\end{equation}
in the presence of the theoretical constraints given in the previous subsection. In addition, we have enforced constraints which can be derived from experimental searches for Supersymmetric charginos and neutralinos at LEP, which 
can be re-interpreted in our model in terms of mass limits on the lightest Higgs states, as follows (see also
Eq.~(\ref{eq:3HDM-exp_constraints}))
$$
        m_{H^{\pm}_{1u}} > 100~{\rm GeV},
$$
\begin{equation}\label{eq:exp_constraints}
         m_{H^0_{1u}}, m_{A^0_{1u}} > m_Z/2.
\end{equation}
(LHC limits do not appear currently to be any stronger \cite{PDG}.) 
%%%%%%%%%%%%%%%%%%%%%%%%%%%%
%For the record, we have generated $10^8$ points, 457170 of which survived all constraints,
%but only  6744 of these yielded a mass spectrum such that one or both of the aforementioned invisible Higgs decays were open
%(i.e.,  $m_h>2m_{H^0_{1u}}$ and/or $m_h>2m_{A^0_{1u}}$).
%%%%%%%%%%%%%%%%%%%%%%%%%%%%
\begin{figure}[!t]
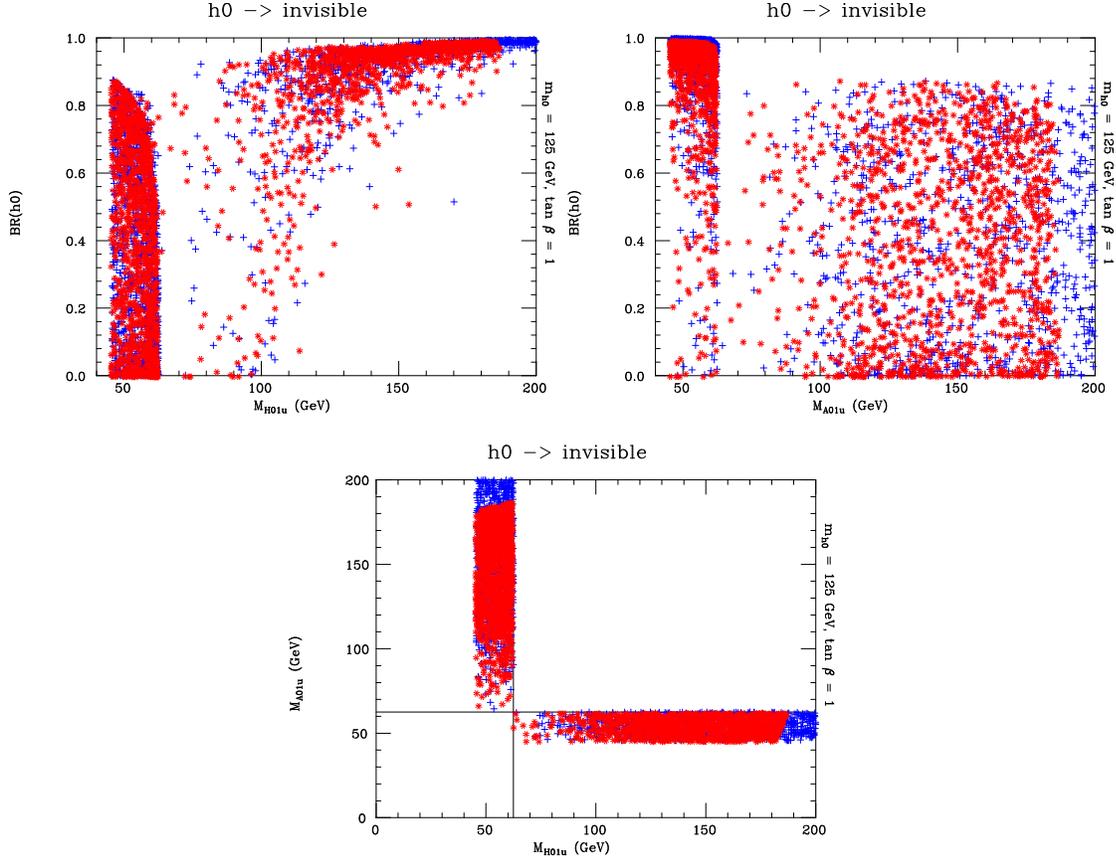

\centering
\includegraphics[width=0.35\linewidth,angle=90]{./tanBeq1-Combined/br_invisible-H01u-combined.ps}
\includegraphics[width=0.35\linewidth,angle=90]{./tanBeq1-Combined/br_invisible-A01u-combined.ps}\\[0.35cm]
\includegraphics[width=0.35\linewidth,angle=90]{./tanBeq1-Combined/DeltaM-combined.ps}
\caption{Parameter scan of the I(4+2)HDM mapped in terms of 
BR$(h\to{H^0_{1u}}{H^0_{1u}})$ + BR$(h\to{A^0_{1u}}{A^0_{1u}})$ as a function of
$m_{H^0_{1u}}$ (top-left) and
$m_{A^0_{1u}}$ (top-right) as well as of the   $m_{H^0_{1u}}$  vs
$m_{A^0_{1u}}$ correlation (bottom)
for $\tan\beta=1$ and $\lambda_{aa}=0(\neq0)$ for the red-star(blue-cross) symbol.
(Recall that the two channels $h\to{H^0_{1u}}{H^0_{1u}}$ and $h\to{A^0_{1u}}{A^0_{1u}}$ are  mutually exclusive,
as explained in the text.)}
\label{fig:tanB=1-mass}
\end{figure}

\begin{figure}[!t]
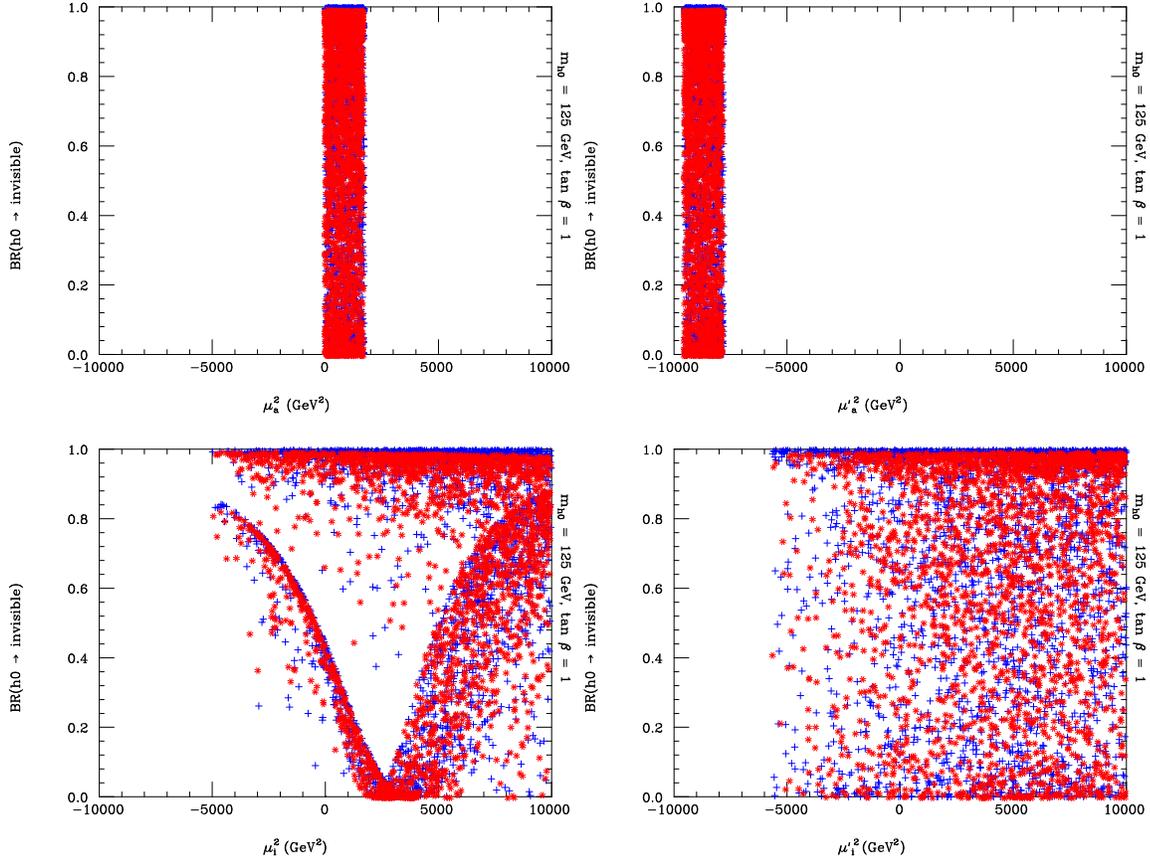

\centering
\includegraphics[width=0.35\linewidth,angle=90]{./tanBeq1-Combined/br_invisible-mua_2-combined.ps}
\includegraphics[width=0.35\linewidth,angle=90]{./tanBeq1-Combined/br_invisible-nua_2-combined.ps}\\[0.35cm]
\includegraphics[width=0.35\linewidth,angle=90]{./tanBeq1-Combined/br_invisible-mui_2-combined.ps}
\includegraphics[width=0.35\linewidth,angle=90]{./tanBeq1-Combined/br_invisible-nui_2-combined.ps}
\caption{Parameter scan of the I(4+2)HDM mapped in terms of 
BR$(h\to{H^0_{1u}}{H^0_{1u}})$ + BR$(h\to{A^0_{1u}}{A^0_{1u}})$ as a function of
${\mu }_a^2$ (top-left),
${\mu'}_a^2$ (top-right),
${\mu }_i^2$ (bottom-left) and
${\mu'}_i^2$ (bottom-right) 
for $\tan\beta=1$ and $\lambda_{aa}=0(\neq0)$ for the red-star(blue-cross) symbol.
(Recall that the two channels $h\to{H^0_{1u}}{H^0_{1u}}$ and $h\to{A^0_{1u}}{A^0_{1u}}$ are mutually exclusive,
as explained in the text.)}
\label{fig:tanB=1-params1}
\end{figure}

\begin{figure}[!t]
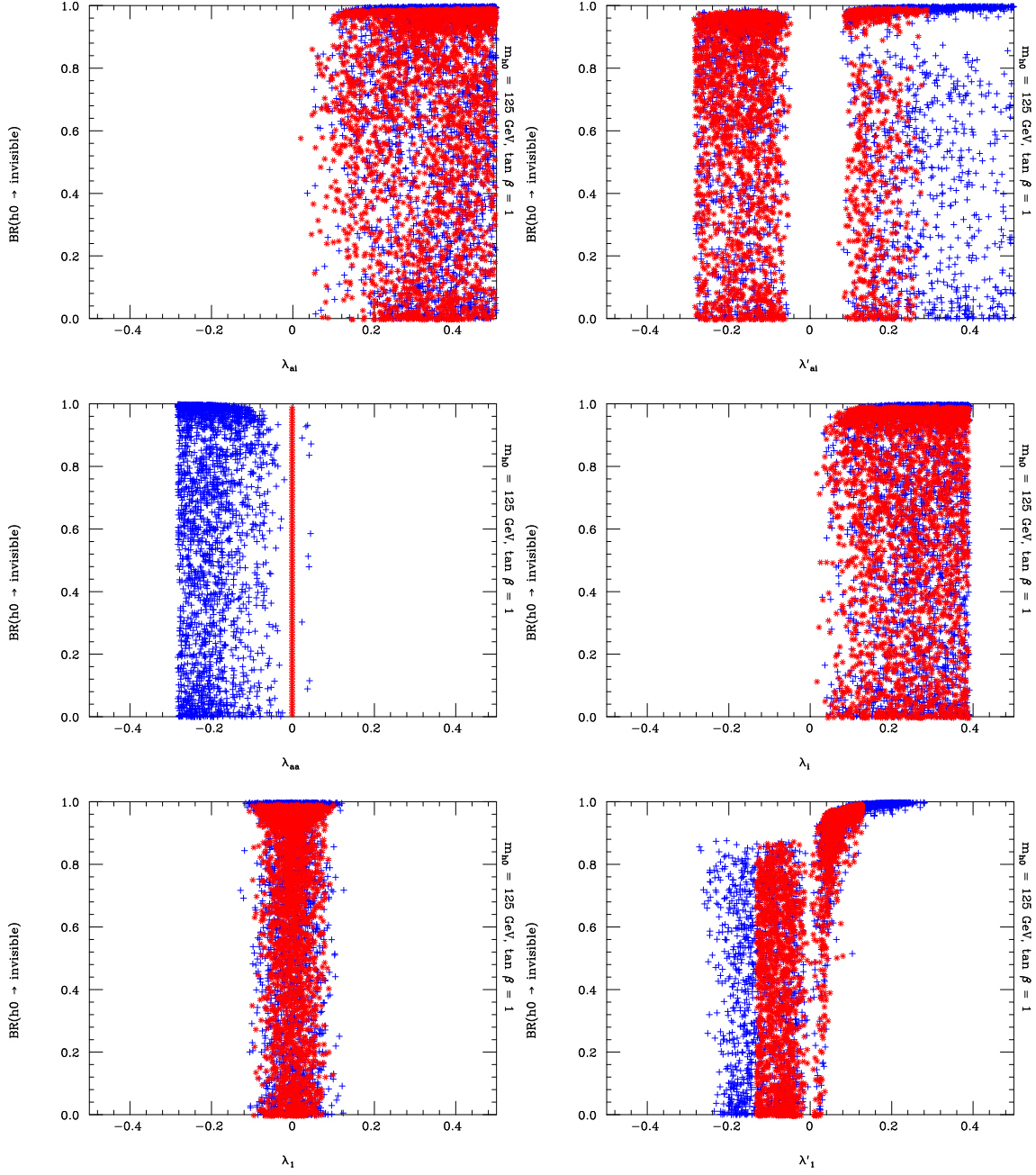

\centering
\includegraphics[width=0.35\linewidth,angle=90]{./tanBeq1-Combined/br_invisible-lamai-combined.ps}
\includegraphics[width=0.35\linewidth,angle=90]{./tanBeq1-Combined/br_invisible-lanai-combined.ps}\\[0.35cm]
\includegraphics[width=0.35\linewidth,angle=90]{./tanBeq1-Combined/br_invisible-lamaa-combined.ps}
\includegraphics[width=0.35\linewidth,angle=90]{./tanBeq1-Combined/br_invisible-lami-combined.ps}\\[0.35cm]
\includegraphics[width=0.35\linewidth,angle=90]{./tanBeq1-Combined/br_invisible-lam1-combined.ps}
\includegraphics[width=0.35\linewidth,angle=90]{./tanBeq1-Combined/br_invisible-lan1-combined.ps}
\caption{Parameter scan of the I(4+2)HDM mapped in terms of 
BR$(h\to{H^0_{1u}}{H^0_{1u}})$ + BR$(h\to{A^0_{1u}}{A^0_{1u}})$ as a function of
${\lambda }_{ai}$ (top-left),
${\lambda'}_{ai}$ (top-right),
${\lambda }_{aa}$ (centre-left),
${\lambda }_{i}$ (centre-right),
${\lambda }_{1}$ (bottom-left) and
${\lambda'}_{1}$ (bottom-right) 
for $\tan\beta=1$ and $\lambda_{aa}=0(\neq0)$ for the red-star(blue-cross) symbol.
(Recall that the two channels $h\to{H^0_{1u}}{H^0_{1u}}$ and $h\to{A^0_{1u}}{A^0_{1u}}$ are  mutually exclusive,
as explained in the text.)}
\label{fig:tanB=1-params2}
\end{figure}

Under the assumption that $\lambda_{aa}=0$, for which we adopt the red-star symbol in the upcoming figures, the top two frames of
Fig.~(\ref{fig:tanB=1-mass}) show the BR of these two decays, as a function
of $m_{H^0_{1u}}$ and $m_{A^0_{1u}}$, separately (recall that they are mutually exclusive, as explained already and exemplified by the bottom frame in
Fig.~(\ref{fig:tanB=1-mass})) whereas  Figs.~(\ref{fig:tanB=1-params1}--\ref{fig:tanB=1-params2})
map the same results over the 9 independent I(4+2)HDM parameters. From the former figure, it is clear that a significant contribution
to the invisible $h$ width emerges in the I(4+2)HDM, which could even be dominant, as the corresponding BR starts from zero and can reach unity (when the decay into ${A^0_{1u}}{A^0_{1u}}$ is the open one). This enables one to 
severely constrain the model, if $\tan\beta=1$, already at present, as current limits from direct searches for invisible $h$ decays place the constraint on BR$(h\to{\rm invisible})$ at less than $\sim40\%$ \cite{invisible-direct} while fits to the LHC Higgs event rates into SM decay modes limit such an observable to no more than 20\% or so  \cite{invisible-indirect,Cheung:2014noa} (as already mentioned). From the latter two figures, we notice that the I(4+2)HDM is not particularly fine tuned to any region of its parameter space, as the spread of surviving points is rather uniform over the tested ranges, with the possible exception
of the $\mu_i^2$ distribution, which sees a cumulation of points towards the edges of the regions delimited by the requirements of positivity of the Higgs mass eigenstates and of the Hessian.    
  
{For the more general, non-zero $\lambda_{aa}$ case we allow it to vary in the usual range
\begin{equation}\label{eq:the_constraints_lamaa}
     -0.5  <  \lambda_{aa}  < 0.5,
\end{equation}
so that one gets the results presented in Figs.~(\ref{fig:tanB=1-mass}--\ref{fig:tanB=1-params2}) using a blue-cross symbol. There is no substantial difference with respect to the previous
case, apart from a noticeable enlargement of the parameter space of the coupling parameters (the $\lambda$'s), owing primarily to the additional degree of freedom.}

In short, even for $\tan\beta=1$, when the couplings of the lightest Higgs boson are exactly those of the SM, the I(4+2)HDM reveals non-SM features which are  testable at the LHC, in the form of  a much larger  
SM-like $h$ invisible decay width, with respect to the SM case, {irrespective of the choice of $\lambda_{aa}$}.

\section{The I(4+2)HDM for general $\tan\beta $}
\label{6HDMtbneq1}
%%%%%%%%%%%%%%%%%%%%%%%%%%%%%%%%%%%%%%%%%%%%%%%%%%%%%%%%%%%%%%%%%%%%%%%%%%%%%%%%%%%%%%%%%%%%%
\subsection{The potential }
As it was shown in section \ref{minimization}, the point $(0,0,0,0,\frac{v_u}{\sqrt{2}},\frac{v_d}{\sqrt{2}})$ is only a minimum of the potential in Eq.~(\ref{simple-potential}) if $\tan\beta = \pm 1$. For general values of $\tan\beta$, $H_{3u}$ and $H_{3d}$ must have non-equal mass terms, $\mu^2_5, \mu^2_6$, in the potential. Therefore, the potential gets the following form:
\bea
\label{general-potential}
V &=& \mu^2_5  |H_{3u}|^2 + \mu^2_6  |H_{3d}|^2  + \mu^2_i \biggl(|H_{1u}|^2 + |H_{1d}|^2 + |H_{2u}|^2 + |H_{2d}|^2 \biggr) \\
&&+ \mu'^2_a \biggl(H^\dagger_{3u} H_{3d}  \biggr) + \mu'^2_i \biggl(H^\dagger_{1d} H_{2u} + H^\dagger_{1d} H_{2d} + H^\dagger_{2u} H_{2d}   \biggr) + h.c. \nonumber\\
&&+ \lambda_a \biggl( |H_{3u}|^2 + |H_{3d}|^2 \biggr)^2 +\lambda_i \biggl( |H_{1u}|^2 + |H_{1d}|^2 + |H_{2u}|^2 + |H_{2d}|^2 \biggr)^2 \nonumber \\
&&+ \lambda_{ai} \biggl[ ( |H_{3u}|^2 + |H_{3d}|^2)(|H_{1u}|^2 + |H_{1d}|^2 + |H_{2u}|^2 + |H_{2d}|^2)   \biggr] \nonumber\\
&&+ \lambda'_{ai} \biggl[|H^\dagger_{3u}H_{1u}|^2+  |H^\dagger_{3u}H_{1d}|^2+ |H^\dagger_{3u}H_{2u}|^2+ |H^\dagger_{3u}H_{2d}|^2 \nonumber\\
&& \qquad \quad + |H^\dagger_{3d}H_{1u}|^2+  |H^\dagger_{3d}H_{1d}|^2+ |H^\dagger_{3d}H_{2u}|^2+ |H^\dagger_{3d}H_{2d}|^2    \biggr] \nonumber\\
&&  {+ \lambda_{aa} \biggl[H^\dagger_{3u}H_{3d} + H^\dagger_{3d}H_{3u} \biggr]^2 } \nonumber\\
&&+ \lambda_{1} \biggl[ H^\dagger_{1d}  H_{3u} + H^\dagger_{1d} H_{3d} + H^\dagger_{2u}H_{3u} + H^\dagger_{2u} H_{3d}+ H^\dagger_{2d}H_{3u} + H^\dagger_{2d}H_{3d} \biggr]^2 +h.c. \nonumber\\
&&+ \lambda'_{1} \biggl[  H^\dagger_{1u} H_{3u}  + H^\dagger_{1u} H_{3d} \biggr]^2 + h.c. \nonumber
\eea
Note that introducing non-equal mass terms for $H_{3u}$ and $H_{3d}$ breaks the $Z^{\phi_5 \leftrightarrow \phi_6}_2$ symmetry and therefore the potential in Eq.~(\ref{general-potential}) is $Z^H_2 \times S_3 \times Z^{DM}_2$-symmetric.

Minimising the potential requires the following equations to be simultaneously satisfied:
\bea 
&& \bullet \quad  v_u \left(\mu^2_5 + \lambda_a (v^2_u+ v^2_d) +{2v^2_d \lambda_{aa}}  \right) + v_d \mu'^2_a = 0 \\[2mm]
&& \bullet \quad v_d \left(\mu^2_6 + \lambda_a (v^2_u+ v^2_d) +{2v^2_u \lambda_{aa}}  \right) + v_u \mu'^2_a = 0 \nonumber
\eea
or equivalently for $ v_u= v' \cos\beta$ and $v_d = v' \sin\beta$, with
\be 
\label{v-general}
 v'^2=  \frac{-\mu^2_5 -  \tan\beta \mu'^2_a}{\lambda_a +2 \sin^2\beta \lambda_{aa}} = 
\frac{-\tan\beta\mu^2_6 - \mu'^2_a}{\tan\beta\lambda_a  +2\cos\beta\sin\beta \lambda_{aa}} 
\ee
where $\tan\beta$ is the solution of the following equation:
\bea 
&& \tan^4\beta \left(\mu'^2_a\lambda_{a} \right) + \tan^3\beta \left((\mu^2_5 - \mu^2_6) \lambda_{a} - 2 \mu^2_6 \lambda_{aa} \right) \nonumber\\
&& + \tan\beta \left((\mu^2_5 - \mu^2_6) \lambda_{a} +2\mu^2_5 \lambda_{aa} \right) +  \left(-\mu'^2_a \lambda_{a} \right) = 0. 
\label{quartic-tanB}
\eea

\subsection{Mass eigenstates }
Upon EWSB the fields acquire the following VEVs:
\be 
\langle H_{1u} \rangle =\langle H_{1d} \rangle =\langle H_{2u}\rangle =\langle H_{2d} \rangle =0, \qquad \langle H_{3u}\rangle = \frac{v_u}{\sqrt{2}}, \quad \langle H_{3d}\rangle =\frac{v_d}{\sqrt{2}}.
\ee
By construction, this pattern of minimum respects the $Z^H_2 \times S_3 \times Z^{DM}_2$ symmetry of the potential. 
 
The study of the potential is simplified if we work in the Higgs basis by rotating the doublets $H_{3u}$ and $H_{3d}$ and defining the new doublets $\widehat{H_{3u}}$ and $\widehat{H_{3d}}$ as
\be  
 \widehat{H_{3u}} = \cos\beta H_{3u} + \sin\beta H_{3d} , \qquad  \widehat{H_{3d}} = -\sin\beta H_{3u} + \cos\beta H_{3d}.
\ee
This rotation changes the VEV alignment to
\bea 
&& \langle \widehat{H_{3u}} \rangle = \cos\beta v_u + \sin\beta v_d = \frac{v'}{\sqrt{2}}, \\
&& \langle \widehat{H_{3d}} \rangle  = -\sin\beta v_u + \cos\beta v_d = 0, \nonumber
\eea
where $ v_u = v'\cos\beta$ and $v_d = v'\sin\beta$ as before. 
%Note that the fields from $\widehat{H_{3u}}$ and $\widehat{H_{3d}}$ are mass eigenstates.

Rewriting the potential in the Higgs basis and expanding it around the point $(0,0,0,0,\frac{v'}{\sqrt{2}},0)$, leads to\footnote{Note that the $v'$ value resulting from Eq.~(\ref{v-general-Higgs-basis}) is equal to the one resulting from Eq.~(\ref{v-general}).}
\be
\label{v-general-Higgs-basis}
v'^2= - \frac{\mu^2_5 +\mu'^2_a +\tan\beta (\mu^2_6 - \mu^2_5 + 2\mu'^2_a) + \tan^2_\beta (\mu^2_6 -\mu'^2_a ) }{\lambda_a (\tan^2_\beta +1 ) + 2\sin\beta\lambda_{aa}(2\sin\beta + \cos\beta -\tan_\beta\sin\beta)} 
\ee
and the following mass spectrum:

\begin{footnotesize}
\bea
\label{mass-general}
&& \textbf{h}\equiv\widehat{\textbf{H}^0_{3u}}  = c_\beta H^0_{3u} + s_\beta H^0_{3d}:
\quad m^2= 
%c^2_\beta \mu^2_5 + s^2_\beta \mu^2_6 + 2c_\beta s_\beta \mu'^2_a + 3\lambda_a v'^2 {+2(1+2c^2_\beta s^2_\beta)\lambda_{aa} v'^2 } = 
2(\lambda_a+\lambda_{aa})v'^2 \\
&& \textbf{H} \equiv \widehat{\textbf{H}^0_{3d}}  = -s_\beta H^0_{3u} + c_\beta H^0_{3d}:
\quad m^2= 
%s^2_\beta \mu^2_5 + c^2_\beta \mu^2_6 - 2c_\beta s_\beta \mu'^2_a + \lambda_a v'^2 {+2(1-2c^2_\beta s^2_\beta)\lambda_{aa} v'^2 } = 
\mu^2_5 + \mu^2_6 +2(\lambda_a+\lambda_{aa})v'^2 \nonumber\\
&& \textbf{G$^\pm$} \equiv \widehat{\textbf{H}^\pm_{3u}}  =c_\beta H^\pm_{3u} + s_\beta H^\pm_{3d}:
\quad m^2= 
%c^2_\beta \mu^2_5 +s^2_\beta \mu^2_6 + 2c_\beta s_\beta \mu'^2_a + \lambda_a v'^2 {+4c^2_\beta s^2_\beta\lambda_{aa} v'^2 } = 
0 \nonumber\\
&& \textbf{H$^\pm$}\equiv \widehat{\textbf{H}^\pm_{3d}} = -s_\beta H^\pm_{3u} + c_\beta H^\pm_{3d}:
\quad m^2= 
%s^2_\beta \mu^2_5 + c^2_\beta \mu^2_6 - 2c_\beta s_\beta \mu'^2_a + \lambda_a v'^2 {-4c^2_\beta s^2_\beta\lambda_{aa} v'^2 } = 
\mu^2_5 + \mu^2_6 +2\lambda_a v'^2 \nonumber\\
&&\textbf{G$^0$}  \equiv \widehat{\textbf{A}^0_{3u}}  = c_\beta A^0_{3u} + s_\beta A^0_{3d}:
\quad m^2= 
%c^2_\beta \mu^2_5 +s^2_\beta \mu^2_6 + 2c_\beta s_\beta \mu'^2_a + \lambda_a v'^2 {+4c^2_\beta s^2_\beta\lambda_{aa} v'^2 } = 
0 \nonumber\\
&&\textbf{A} \equiv \widehat{\textbf{A}^0_{3d}}  =-s_\beta A^0_{3u} + c_\beta A^0_{3d}:
\quad m^2=
%s^2_\beta \mu^2_5 + c^2_\beta \mu^2_6 - 2c_\beta s_\beta \mu'^2_a + \lambda_a v'^2 {-4c^2_\beta s^2_\beta\lambda_{aa} v'^2 }= 
\mu^2_5 + \mu^2_6 +2\lambda_a v'^2 \nonumber\\
&& \textbf{H$^{\pm}_{1u}$}  : 
\quad m^2={\mu_i}^2+\frac{\lambda_{ai}}{2} v'^2 \nonumber\\[2mm]
&& \textbf{H$^0_{1u}$} : 
\quad m^2=\mu^2_i + (\frac{\lambda_{ai}+ \lambda'_{ai}}{2} + (1+2c_\beta s_\beta)\lambda'_1  ) v'^2 \nonumber\\[2mm]
&& \textbf{A$^0_{1u}$} : 
\quad m^2=\mu^2_i + (\frac{\lambda_{ai}+ \lambda'_{ai}}{2} - (1+2c_\beta s_\beta)\lambda'_1 ) v'^2 \nonumber\\[2mm]
&& \textbf{H$_x$} = \frac{H^0_{1d}-H^0_{2d}}{\sqrt{2}}: 
\quad m^2=\mu^2_i- \mu'^2_i + (\frac{\lambda_{ai}+ \lambda'_{ai}}{2})v'^2 \nonumber\\[2mm]
&& \textbf{H$_y$}=\frac{H^0_{1d}-H^0_{2u}}{\sqrt{2}}: 
\quad m^2=\mu^2_i- \mu'^2_i + (\frac{\lambda_{ai}+ \lambda'_{ai}}{2})v'^2 \nonumber\\[2mm]
&& \textbf{H$_z$} =\frac{H^0_{1d}+H^0_{2u}+H^0_{2d}}{\sqrt{3}}:  
\quad m^2=\mu^2_i+ 2\mu'^2_i + (\frac{\lambda_{ai}+ \lambda'_{ai}}{2}+ 3(1+2c_\beta s_\beta)\lambda_1 )v'^2 \nonumber\\
&& \textbf{A$_x$}=\frac{A^0_{1d}-A^0_{2d}}{\sqrt{2}}: 
\quad m^2=\mu^2_i- \mu'^2_i + (\frac{\lambda_{ai}+ \lambda'_{ai}}{2})v'^2 \nonumber\\[2mm]
&& \textbf{A$_y$}=\frac{A^0_{1d}-A^0_{2u}}{\sqrt{2}}: 
\quad m^2=\mu^2_i- \mu'^2_i + (\frac{\lambda_{ai}+ \lambda'_{ai}}{2})v'^2 \nonumber\\[2mm]
&& \textbf{A$_z$}=\frac{A^0_{1d}+A^0_{2u}+A^0_{2d}}{\sqrt{3}}: 
\quad m^2= \mu^2_i+ 2\mu'^2_i + (\frac{\lambda_{ai}+ \lambda'_{ai}}{2} -3(1+2c_\beta s_\beta)\lambda_1 )v'^2 \nonumber\\
&& \textbf{H$^\pm_x$} = \frac{H^{\pm}_{1d}-H^{\pm}_{2d}}{\sqrt{2}}: 
\quad m^2=\mu^2_i- \mu'^2_i + \frac{\lambda_{ai}}{2}v'^2 \nonumber\\[2mm]
&& \textbf{H$^\pm_y$} =\frac{H^{\pm}_{1d}-H^{\pm}_{2u}}{\sqrt{2}}: 
\quad m^2= \mu^2_i- \mu'^2_i + \frac{\lambda_{ai}}{2}v'^2 \nonumber\\[2mm]
&& \textbf{H$^\pm_z$} =\frac{H^{\pm}_{1d}+H^{\pm}_{2u}+H^{\pm}_{2d}}{\sqrt{3}}: 
\quad  m^2=\mu^2_i+ 2\mu'^2_i + \frac{\lambda_{ai}}{2}v'^2 \nonumber
\eea
\end{footnotesize}
where $c_\beta$ and $s_\beta$ are $\cos\beta$ and $\sin\beta$ respectively.
The Feynman rules for this potential are presented in Appendix \ref{general}.

%\begin{center}
%\includegraphics[scale=1.1]{Masses-6HDM-tanbnot1.ps} 
%\end{center}

\subsection{Constraints on the parameters}
As it will be shown from the Feynman rules, for $\tan \beta \neq 1$, again, only the state $h$ can couple to $W^\pm$ and $Z$ gauge bosons and play the role of the 
125 GeV SM-like Higgs. Here, the condition $m^2_{h} < m^2_{H}$ leads to
\be 
\bullet \quad \mu^2_5 + \mu^2_6 > 0
\ee

We also require the neutral DM candidates decay channel to the charged field from the $H_{1u}$ doublet to be closed, therefore:
%\begin{footnotesize}
\begin{itemize}
\item {if $H^0_{1u}$ is the DM candidate: } 
\bea 
&& m^2_{H^0_{1u}} < 2 m^2_{H^\pm_{1u}} \quad \rightarrow \quad   ((1+2c_\beta s_\beta) \lambda'_1 + \frac{1}{2} \lambda'_{ai})v'^2  < \mu^2_i +  \frac{1}{2} \lambda_{ai}v'^2  \nonumber
\eea
\item {if $A^0_{1u}$ is the DM candidate: } 
\bea
&& m^2_{A^0_{1u}} < 2 m^2_{H^\pm_{1u}} \quad \rightarrow \quad ( - (1+2c_\beta s_\beta) \lambda'_1 + \frac{1}{2} \lambda'_{ai})v'^2  < \mu^2_i +  \frac{1}{2} \lambda_{ai}v'^2 \nonumber
\eea
\end{itemize}
%\end{footnotesize}

Further constraints on the parameters are required by:
\begin{enumerate}
\item 
\textbf{Positivity of the mass eigenstates}:
\bea
&& \bullet \quad \lambda_a +\lambda_{aa} > 0    \\
&& \bullet \quad \mu^2_5 + \mu^2_6 + 2\lambda_a v'^2  > 0  \nonumber\\
&& \bullet \quad \mu^2_5 + \mu^2_6 + 2(\lambda_a + \lambda_{aa}) v'^2  > 0  \nonumber\\
%&& \bullet \quad \mu^2_i + \frac{\lambda_{ai}}{2} v'^2 > 2|\mu'^2_i| \nonumber\\
&& \bullet \quad \mu^2_i + \frac{\lambda_{ai}}{2} v'^2 > \mu'^2_i \quad \mbox{for} \quad \mu'^2_i>0 \nonumber\\
&& \bullet \quad \mu^2_i + \frac{\lambda_{ai}}{2} v'^2 > 2|\mu'^2_i| \quad \mbox{for} \quad \mu'^2_i<0 \nonumber\\
&& \bullet \quad \mu^2_i + \frac{\lambda_{ai}+ \lambda'_{ai}}{2} v'^2  > \mu'^2_i \nonumber\\
&& \bullet \quad \mu^2_i + \frac{\lambda_{ai}+ \lambda'_{ai}}{2} v'^2  >   |(1+ 2c_\beta s_\beta)\lambda'_1 | v'^2 \nonumber\\
&& \bullet \quad \mu^2_i +2\mu'^2_i + \frac{\lambda_{ai}+ \lambda'_{ai}}{2} v'^2  >  3 |(1+ 2c_\beta s_\beta)\lambda_1 | v'^2 \nonumber
\eea

\item
\textbf{Bounded-ness of the potential}:
\bea
&& \bullet \quad  \lambda_{a} > 0 \\
&& \bullet \quad  \lambda_i   >0    \nonumber\\
&& \bullet \quad  \lambda_{ai} + \lambda'_{ai} > -2 \sqrt{\lambda_{a}\lambda_{i}} \nonumber\\
&& {\bullet \quad 4\lambda_a + \lambda_{aa} > 0} \nonumber
\eea
and we require the parameters of the $V_{Z_2}$ part to be smaller than the parameters of the $V_0$ part of the potential:
\be 
 \bullet \quad |\lambda_1|, |\lambda'_1| < |\lambda_a|, |\lambda_i|, |\lambda_{ai}|, |\lambda'_{ai}|, {|\lambda_{aa}| }
\ee

\item
\textbf{Positive-definite-ness of the Hessian}:
\bea
&& \bullet \quad \mu^2_i + \frac{\lambda_{ai}+ \lambda'_{ai}}{2} v'^2  > 0 \\
&& \bullet \quad \mu^2_i + \frac{\lambda_{ai}+ \lambda'_{ai}}{2} v'^2  > \mu'^2_i \nonumber\\
&& \bullet \quad \mu^2_i + \frac{\lambda_{ai}+ \lambda'_{ai}}{2} v'^2  > -2\mu'^2_i \nonumber\\
&& {\bullet \quad \biggl(\mu^2_5 + \mu^2_6  + (3\lambda_a + 2\lambda_{aa}) v'^2 \biggr)^2 > 
4 \biggl( \mu'^2_a + ( \lambda_a + 6\lambda_{aa})v'^2 s_\beta c_\beta \biggr)^2} \nonumber\\
&& {\quad + \biggl(\mu^2_5 -\mu^2_6 + (\lambda_a - 2\lambda_{aa}) v'^2(c^2_\beta -s^2_\beta)  \biggr)^2 } \nonumber
\eea 

\end{enumerate}

\subsection{Results for the general I(4+2)HDM case}

In the $\tan\beta\neq1$ case the parameter $\mu_a$ is replaced by $\mu_5$ and $\mu_6$, which are scanned
over the same interval, i.e.,
$$
        0 < {\mu}_5^2, \mu_6^2 < 10~{\rm TeV}^2.  
$$
It is worth noting that in this case both $v'$ and $\tan\beta$ are derived parameters. The latter, in particular, is obtained
through a quartic equation (Eq.~(\ref{quartic-tanB})). We find that only two sets of solutions are possible:
\begin{enumerate}
\item two real roots and two complex conjugate roots;
\item four roots real and distinct.
\end{enumerate}
In each case we only keep the real roots and amongst these we select only one randomly for each set of input parameters.
However, not all $\tan\beta$ values obtained this way are phenomenologically viable. We restrict these over
the interval
$$
1 < \tan\beta < 50,
$$   
so as to not induce unacceptable Yukawa couplings for the I(4+2)HDM considered. Also, contrary to the case $\tan\beta=1$, here, experimental 
constraints extracted from the LHC Higgs search are no longer automatically satisfied, owing to the Yukawa couplings
being different from the SM case. Hence, we have proceeded as follows. We have asked the $h$ state to have a mass in a window 10 GeV wide centred
around 125 GeV (recall that now $m_h$ is a derived quantity) and, at the same time, its coupling strengths to fall within the error bands of either the ATLAS or CMS 
(when not both) measurements. The experimental results adopted here from ATLAS~\cite{ATLAS:2012au,Aad:2012tfa,ATLASNOTE2:2013,ATLAS-CONF-2013-072} and CMS~\cite{Chatrchyan:2012ufa, :2013tq, Chasco:2013pwa, CMS-PAS-HIG-13-005}  are
based on an integrated luminosity of
$4.7$ fb$^{-1}$ at $\sqrt{s} =7$ TeV plus $13$ fb$^{-1}$ at $\sqrt{s}=8$ TeV (ATLAS) and $5.1$ fb$^{-1}$ at $\sqrt{s} =7$ TeV plus $19.6$ fb$^{-1}$ at $\sqrt{s}=8$ TeV (CMS). 
The results reported by ATLAS for the signal strengths are given by \cite{ATLAS:2012au,Aad:2012tfa,ATLASNOTE2:2013,ATLAS-CONF-2013-072}:%
\bea%
\mu_{b\bar b} &=& -0.4 \pm 1.0,\nonumber\\
\mu_{\tau^+\tau^-} &=& 0.8 \pm 0.7,\nonumber\\
\mu_{\gamma\gamma} &=& 1.6 \pm 0.3,\nonumber\\
\mu_{WW} &=& 1.0 \pm 0.3,\nonumber\\
\mu_{ZZ} &=& 1.5 \pm 0.4. 
\eea%
From the CMS collaboration one has instead \cite{Chatrchyan:2012ufa, :2013tq, Chasco:2013pwa, CMS-PAS-HIG-13-005}:
\bea%
\mu_{b\bar b} &=& 1.15 \pm 0.62,\nonumber\\
\mu_{\tau^+\tau^-} &=& 1.1 \pm 0.41,\nonumber\\
\mu_{\gamma\gamma} &=& 0.77 \pm 0.27,\nonumber\\
\mu_{_{WW}} &=& 0.68 \pm 0.20,\nonumber\\
\mu_{_{ZZ}} &=&  0.92 \pm 0.28.
\eea%
(Different measurements may
be found in the specialized literature, however, we have verified that our results are rather insensitive to the consequent variations, including the recent revision by the CMS
collaboration of the di-photon signal strength \cite{CMS-AA-latest}.) All other experimental constraints enforced are as before: see Eq.~(\ref{eq:exp_constraints}). 

\begin{figure}[!t]
\centering
\includegraphics[width=0.35\linewidth,angle=90]{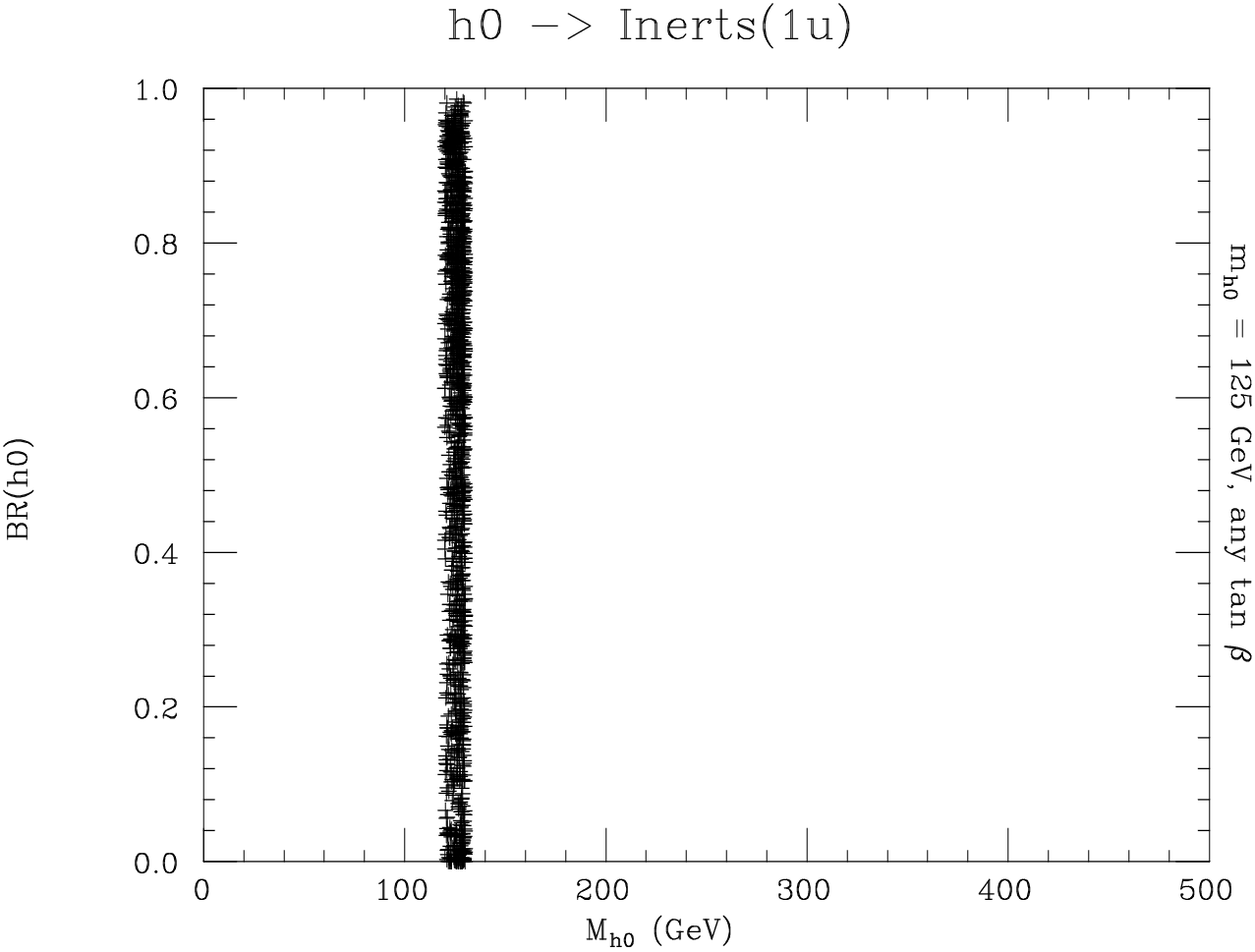}
\includegraphics[width=0.35\linewidth,angle=90]{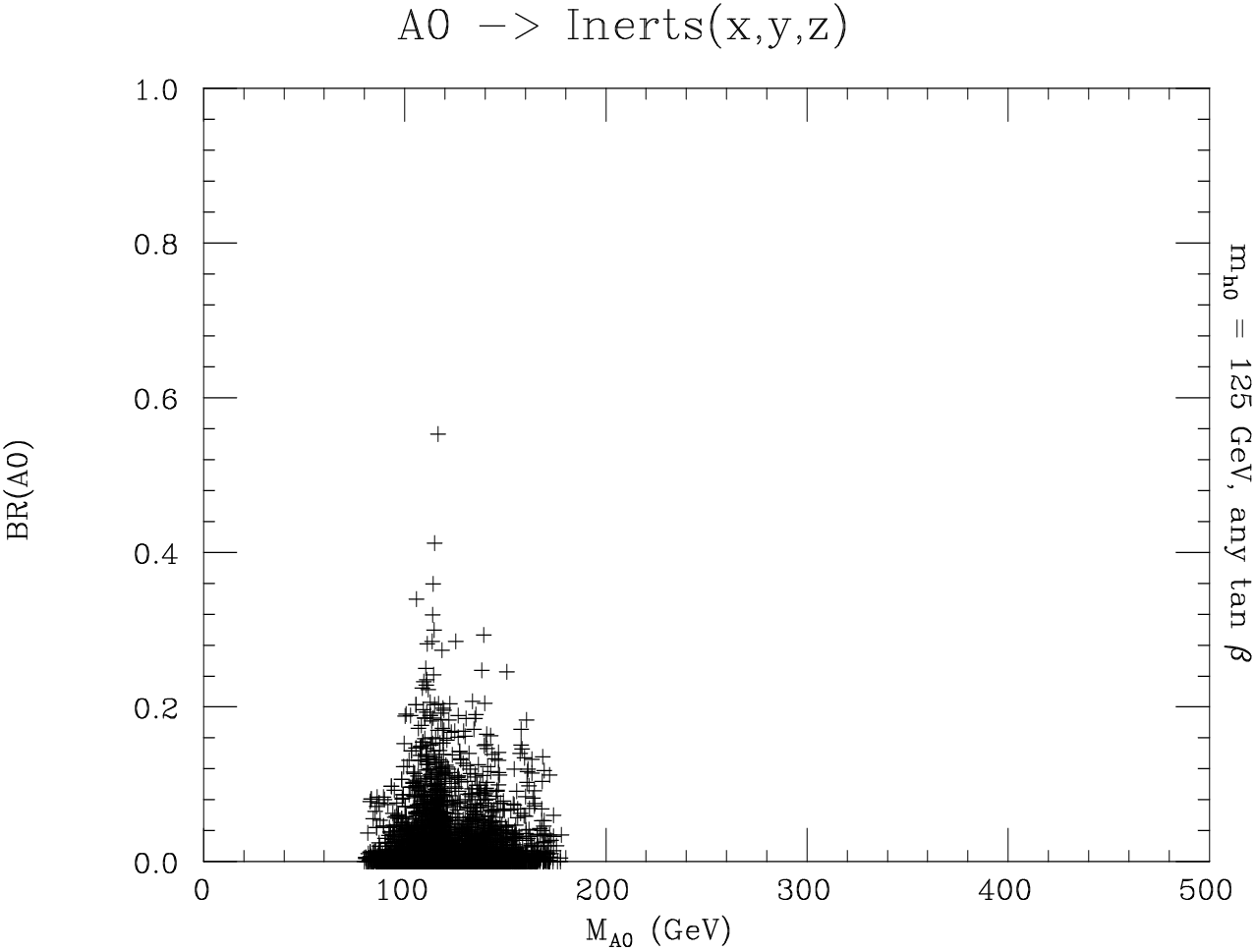}\\[1.0cm]
\includegraphics[width=0.35\linewidth,angle=90]{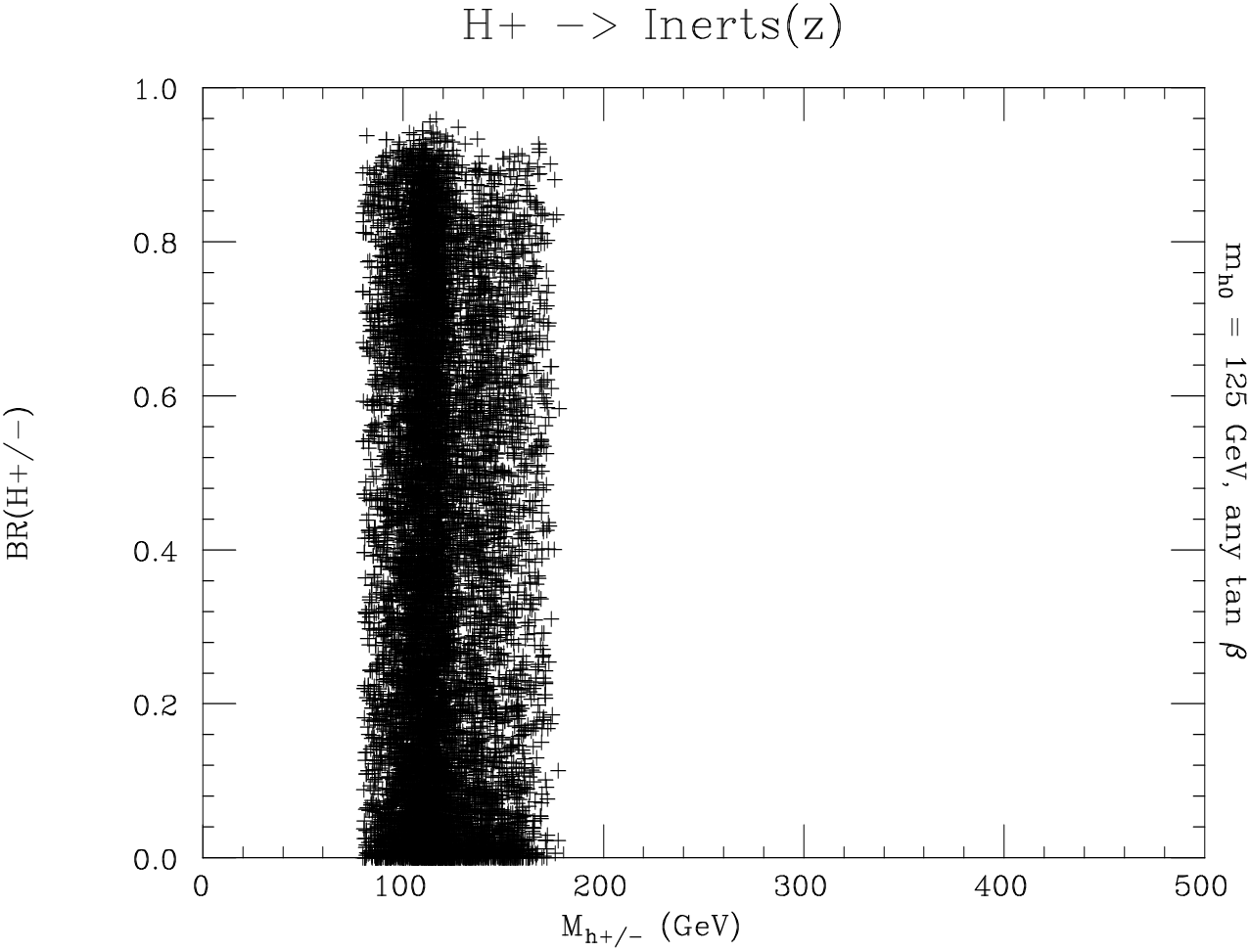}
\caption{Parameter scan of the I(4+2)HDM mapped in terms of 
BR$(h\to{H^0_{1u}}{H^0_{1u}})$ + BR$(h\to{A^0_{1u}}{A^0_{1u}})$ (top-left),
BR$(A\to{A_{x}}{H_{x}})$ +
BR$(A\to{A_{y}}{H_{y}})$ + 
BR$(A\to{A_{x}}{H_{y}})$ + 
BR$(A\to{A_{y}}{H_{x}})$  +
BR$(A\to{A_{z}}{H_{z}})$   (top-right) and
BR$(H^\pm\to{H^\pm_{z}}{H_{z}})$ + BR$(H^\pm\to{H^\pm_{z}}{A_{z}})$ (bottom)
as a function of the mass of the decaying Higgs state
for $\tan\beta\neq1$ and $\lambda_{aa}\ne0$.}
\label{fig:tanB=/=1-Decays-split}
\end{figure}

The interesting non-SM-like Higgs decay channels which emerge following our constrained scan are the ones below\footnote{Notice that, depending on the values of the initial and final state Higgs masses involved, not all the subchannels might be open at the same time. Notice also that even though the $H\to hh$ vertex appears in the Feynman rules, $H$ is never heavy enough to have $m_{H}> 2m_h=250$ GeV, which is line with the $A$ and $H^\pm$ mass spectra singled out by the scan (see Fig.(\ref{fig:tanB=/=1-Decays-split})).}:
\begin{enumerate}
\item
BR$(h\to{H^0_{1u}}{H^0_{1u}})$ + BR$(h\to{A^0_{1u}}{A^0_{1u}})$,
\item
BR$(A\to{A_{x}}{H_{x}})$ +
BR$(A\to{A_{y}}{H_{y}})$ + 
BR$(A\to{A_{x}}{H_{y}})$ + 
BR$(A\to{A_{y}}{H_{x}})$  +
BR$(A\to{A_{z}}{H_{z}})$,
\item
BR$(H^\pm\to{H^\pm_{z}}{H_{z}})$ + BR$(H^\pm\to{H^\pm_{z}}{A_{z}})$.
\end{enumerate}
Their distributions as a function of the mass of the Higgs state which is decaying are found in Fig.~(\ref{fig:tanB=/=1-Decays-split}).
All  such channels can be sizeable, particularly the first and last one, which can in fact be dominant (100\% and 95\%, respectively),
whereas the second one can (exceptionally) reach the 60\% level. 
These are Higgs boson decays which are very distinctive of this model. The first one yields the usual
invisible Higgs signature, with the two channels $h\to{H^0_{1u}}{H^0_{1u}}$ and $h\to{A^0_{1u}}{A^0_{1u}}$ being mutually exclusive (as seen already for the
$\tan\beta=1$ case),
indeed compatible with current data (as previously discussed).  
The second and third decay, which are relevant over a common interval in mass for the CP-odd and charged Higgs boson (between   80 and 180 GeV), are particularly interesting, as they lead to {Higgs-to-two-Higgs or Higgs-to-gauge+Higgs 
 (cascade) decays
in the inert Higgs sector, since the heavy inert Higgs particles appearing in the final state are unstable and will eventually decay into the DM candidates, in the absence of the $Z^{DM}_2$ symmetry. Note that the heavy inert CP-even particles $(H_x,H_y,H_z)$ have the same $Z_2^{H}$ charge as the DM candidate. However, under $Z_2^{DM}$ the $(H_x,H_y,H_z)$ particles have a different quantum number to $H^0_{1u}$ and therefore do not decay to $H^0_{1u}$. They have an $S_3$ symmetry among themselves and provide another stable DM sector. This is the same scenario as in multi-component DM models.} 
{Hence, in the absence of $Z^{DM}_2$, all inert particles carry the same $Z_2^{H}$ charge, and therefore the heavy inert particles $(H_x,H_y,H_z)$ decay to the lightest inert particle, $H^0_{1u}$, which is the DM candidate, resulting in the aforementioned cascade decays which are of novel phenomenological interest.} As the computation of the inert Higgs decay rates, which would be needed to estimate the probability of the various cascade patterns, was beyond the scope of this paper, we defer its study to a future publication.  Here, we limit ourselves to confirm that the mass spectra generated in the inert sector of the I(4+2)HDM do enable such cascade patterns.

%%%%%%%%%%%%%%%%%%%%%%%%%%%%%%%%%%%%%%%%%%%%%%%%%%%%%%%%%%%%%%%%%%%%%%%%%%%%%%%%%%%%%%%%%%%%%

\section{Conclusion}
\label{conclusion}

We have made a phenomenological study of a model
with two inert doublets plus one Higgs doublet (I(2+1)HDM)
which is symmetric under a $Z_2$ group, preserved after EWSB by the vacuum alignment $(0,0,v)$. This model may be regarded as an extension of the I(1+1)HDM, by the addition of an extra inert scalar doublet. 
The doublets are termed ``inert'' since they do not develop a VEV, nor do they couple to fermions.
The lightest neutral field from the two inert doublets provides a 
viable DM candidate, which we labelled $H_1$ which is stabilised by the conserved $Z_2$ symmetry. 

We have studied the new Higgs decay channels offered by the scalar fields from the extra doublets and their effect on the SM Higgs couplings. The main signature of such models is that of invisible Higgs decays from $h\to{H_{1}}{H_{1}}$. However we have also identified and studied a new decay channel into off-shell photon(s), originating EM showers of up to about 30 GeV plus missing energy, which distinguishes the I(2+1)HDM from the I(1+1)HDM. 

Motivated by Supersymmetry, which requires an even number of doublets,
we then extended this model into another with four inert doublets plus one Higgs doublet (I(4+2)HDM), for example, as in the E$_6$SSM \cite{King:2005jy,King:2005my}.
The first inert doublet $\phi_1$ is odd under a $Z^{DM}_2$ symmetry leading to the
lightest $\phi_1$ state being stable, while the 
three inert doublets, $\phi_2$, $\phi_3$ and $\phi_4$, are also protected by an unbroken $Z^H_2$ symmetry and the lightest state from this sector will also be absolutely stable.

We have analysed the theory and phenomenology of the above I(4+2)HDM for the special 
$\tan\beta=1$ case, before turning to the general $\tan\beta \neq 1$ case. In the $\tan\beta=1$ case the lightest neutral state from $\phi_1$, either $H^0_{1u}$ or $A^0_{1u}$, is absolutely stable. In this case the invisible SM-like Higgs decay width into $H^0_{1u}H^0_{1u}$ or $A^0_{1u}A^0_{1u}$ can be rather large.

Similarly for $\tan\beta \neq 1$, the lighter state between $H^0_{1u}$ and $A^0_{1u}$ is absolutely stable.
In this case the SM-like Higgs decays into 
$H^0_{1u}H^0_{1u}$ and $A^0_{1u}A^0_{1u}$ can either 
involve invisible Higgs decays, or a final state where 
the heavier of $H^0_{1u}$ and $A^0_{1u}$ will annihilate into the lighter one.
Finally, for $\tan\beta \neq 1$, 
we also have considered some rather general possibilities for new Higgs physics
in which the CP-odd state
$A$ and charged Higgs $H^{\pm}$ may decay into heavier inert states, leading to a rich 
physics of cascade decays which deserves further investigation.

In conclusion, following the discovery of a SM-like Higgs boson, it is entirely possible that 
further Higgs bosons await discovery at the LHC. It is also possible that 
one or more of these new Higgs bosons may be responsible for the DM of the universe.
Such considerations motivate the Inert class of NHDMs considered here and, if such models
are realised in Nature, then a novel and very rich phenomenology will be in store for the LHC, as 
our study has shown.

\section{Acknowledgements}
VK acknowledges numerous useful discussions with Andrew G. Akeroyd. Part of VK's research was financed through a Visiting Fellowship from The Leverhulme Trust (London, UK). SM is financed in part through the NExT Institute and from the STFC Consolidated ST/J000396/1. SFK also acknowledges partial support from the STFC Consolidated ST/J000396/1 and EU ITN grant INVISIBLES 289442.

\appendix

\section{Feynman rules in the I(2+1)HDM}\label{Feyn-3HDM}

\subsection{Scalar couplings}

\bea
&& hh \longrightarrow h \qquad \qquad  \lambda_{33} v \nonumber\\ 
&& H^+_1 H^-_1 \longrightarrow h  \qquad (\lambda_{23} \cos^2\theta_c +\lambda_{31}\sin^2\theta_c) v \nonumber\\
&& H^+_2 H^-_2 \longrightarrow h  \qquad (\lambda_{23}\sin^2\theta_c+\lambda_{31}\cos^2\theta_c) v  \nonumber\\ 
&& H^\pm_2 H^\mp_1 \longrightarrow h  \qquad (\lambda_{23}-\lambda_{31})\sin\theta_c \cos\theta_c v \nonumber\\ 
&& H_1 H_1 \longrightarrow h \qquad  ((\lambda_{23}+ \lambda'_{23}+2\lambda_{2})\cos^2\theta_h +( \lambda_{31} +\lambda'_{31}+2\lambda_{3})\sin^2\theta_h) \frac{v}{2} \nonumber\\
&& H_2 H_2 \longrightarrow h \qquad  ((\lambda_{23}+ \lambda'_{23}+2\lambda_{2})\sin^2\theta_h  +( \lambda_{31} +\lambda'_{31}+2\lambda_{3})\cos^2\theta_h)\frac{v}{2} \nonumber\\ 
&& H_2 H_1 \longrightarrow h  \qquad (\lambda_{23}+\lambda'_{23}+2\lambda_{2}  -(\lambda_{31}+\lambda'_{31}+2\lambda_{3})) \sin\theta_h \cos\theta_h v \nonumber\\
&& A_1 A_1 \longrightarrow h  \qquad ((\lambda_{23}+ \lambda'_{23}-2\lambda_{2})\cos^2\theta_a +( \lambda_{31} +\lambda'_{31}-2\lambda_{3})\sin^2\theta_a) \frac{v}{2}  \nonumber\\
&& A_2 A_2 \longrightarrow h  \qquad ((\lambda_{23}+ \lambda'_{23}-2\lambda_{2})\sin^2\theta_a  +( \lambda_{31} +\lambda'_{31}-2\lambda_{3})\cos^2\theta_a) \frac{v}{2} \nonumber\\
&& A_2 A_1 \longrightarrow h  \qquad (\lambda_{23}+\lambda'_{23}+2\lambda_{2}  -(\lambda_{31}+\lambda'_{31}+2\lambda_{3})) \sin\theta_a \cos\theta_a v  \nonumber
\eea

\subsection{Gauge couplings}
\bea
&& W^+ W^- \longrightarrow h \qquad \qquad \frac{g^2}{2} v \nonumber\\[2mm]
&& ZZ \longrightarrow h \qquad \qquad \frac{1}{8}(g\cos\theta_W + g'\sin\theta_W)^2 v \nonumber\\[2mm]
&& H^+_1 H^-_1, H^+_2 H^-_2 \longrightarrow \gamma \qquad \qquad \frac{i}{2}(g\sin\theta_W + g'\cos\theta_W)(K+K')^\mu \nonumber\\[2mm]
%&& H^\pm_1 H^\mp_2 \longrightarrow \gamma \qquad \qquad -\frac{i}{2}(g\sin\theta_W + g'\cos\theta_W)q'p'(1+a'b')(K+K')^\mu = 0 \nonumber\\[2mm]
%
&& H^+_1 H^-_1, H^+_2 H^-_2 \longrightarrow Z \qquad \qquad \frac{i}{2}(g\cos\theta_W - g'\sin\theta_W)(K+K')^\mu \nonumber\\[2mm]
%&& H^\pm_1 H^\mp_2 \longrightarrow Z \qquad \qquad -\frac{i}{2}(g\cos\theta_W - g'\sin\theta_W)q'p'(1+a'b')(K+K')^\mu = 0 \nonumber\\[2mm]
%
&& H^\pm_2 H_2, H^\pm_1 H_1 \longrightarrow W^\pm \qquad \qquad \frac{ig}{2} \cos(\theta_h -\theta_c)(K+K')^\mu \nonumber\\[2mm]
&& H^\pm_2 H_1, H^\pm_1 H_2 \longrightarrow W^\pm \qquad \qquad \frac{ig}{2} \sin(\theta_h -\theta_c)(K+K')^\mu \nonumber\\[2mm]
&& H^\pm_2 A_2, H^\pm_1 A_1 \longrightarrow W^\pm \qquad \qquad  \frac{g}{2} \cos(\theta_a -\theta_c)(K+K')^\mu \nonumber\\[2mm]
&& H^\pm_2 A_1, H^\pm_1 A_2 \longrightarrow W^\pm \qquad \qquad  \frac{g}{2} \sin(\theta_a -\theta_c)(K+K')^\mu \nonumber\\[2mm]
&& H_2 A_2, H_1 A_1 \longrightarrow Z \qquad \qquad   \frac{1}{2}(g\cos\theta_W + g'\sin\theta_W) \cos(\theta_h -\theta_a)(K+K')^\mu \nonumber\\[2mm]
&& H_2 A_1, H_1 A_2 \longrightarrow Z \qquad \qquad  \frac{1}{2}(g\cos\theta_W + g'\sin\theta_W) \sin(\theta_h -\theta_a)(K+K')^\mu \nonumber
\eea
where $g$ and $g'$ are the coupling constants associated with the groups $SU(2)_L$ and $U(1)_Y$ respectively, $\theta_W$ is the Weinberg angle, $K$ and $K'$ are the momenta of the associated particles.

The Yukawa couplings in the 3HDM case with $(0,0,\frac{v}{2})$ are identical to the SM ones.

\section{Feynman rules in the I(4+2)HDM with $\tan\beta=1$}\label{simple}

\subsection{Scalar couplings}
\bea
&& H^+ H^- \longrightarrow h \qquad \qquad 2\sqrt{2}\lambda_a v { -2\sqrt{2}\lambda_{aa} v} \nonumber\\
&& hh \longrightarrow h \qquad \qquad \sqrt{2}\lambda_a v { +\sqrt{2}\lambda_{aa} v} \nonumber\\
&& HH \longrightarrow h \qquad \qquad \sqrt{2}\lambda_a v { -\sqrt{2}\lambda_{aa} v}\nonumber\\
&& AA \longrightarrow h \qquad \qquad \sqrt{2}\lambda_a v { -\sqrt{2}\lambda_{aa} v}\nonumber\\
&& H^+_{1u} H^-_{1u} \longrightarrow h \qquad \qquad \sqrt{2}\lambda_{ai} v \nonumber\\
&& H^+_{z} H^-_{z} \longrightarrow h \qquad \qquad \sqrt{2}\lambda_{ai} v \nonumber\\
&& H^+_{x} H^-_{x} \longrightarrow h \qquad \qquad \frac{4\sqrt{2}}{3}\lambda_{ai} v \nonumber\\
&& H^+_{y} H^-_{y} \longrightarrow h \qquad \qquad \frac{4\sqrt{2}}{3}\lambda_{ai} v \nonumber\\
&& H^\pm_{x} H^\mp_{y} \longrightarrow h \qquad \qquad -\frac{2\sqrt{2}}{3}\lambda_{ai} v \nonumber\\
&& H^0_{1u} H^0_{1u} \longrightarrow h \qquad \qquad \frac{\sqrt{2}}{2}(\lambda_{ai}+\lambda'_{ai})v + 2\sqrt{2}\lambda'_1 v \nonumber\\
&& A^0_{1u} A^0_{1u} \longrightarrow h \qquad \qquad \frac{\sqrt{2}}{2}(\lambda_{ai}+\lambda'_{ai})v - 2\sqrt{2}\lambda'_1 v \nonumber\\
&& H_z H_z \longrightarrow h \qquad \qquad \frac{\sqrt{2}}{2}(\lambda_{ai}+\lambda'_{ai})v + 6\sqrt{2}\lambda_1 v \nonumber\\
&& A_z A_z \longrightarrow h \qquad \qquad \frac{\sqrt{2}}{2}(\lambda_{ai}+\lambda'_{ai})v - 6\sqrt{2}\lambda_1 v \nonumber\\
&& H_x H_x \longrightarrow h \qquad \qquad \frac{2\sqrt{2}}{3}(\lambda_{ai}+\lambda'_{ai})v  \nonumber\\
&& A_x A_x \longrightarrow h \qquad \qquad \frac{2\sqrt{2}}{3}(\lambda_{ai}+\lambda'_{ai})v  \nonumber\\
&& H_y H_y \longrightarrow h \qquad \qquad \frac{2\sqrt{2}}{3}(\lambda_{ai}+\lambda'_{ai})v  \nonumber\\
&& A_y A_y \longrightarrow h \qquad \qquad \frac{2\sqrt{2}}{3}(\lambda_{ai}+\lambda'_{ai})v  \nonumber\\
&& H_x H_y \longrightarrow h \qquad \qquad -\frac{2\sqrt{2}}{3}(\lambda_{ai}+\lambda'_{ai})v  \nonumber\\
&& A_x A_y \longrightarrow h \qquad \qquad -\frac{2\sqrt{2}}{3}(\lambda_{ai}+\lambda'_{ai})v  \nonumber
\eea

\subsection{Gauge couplings}\label{gauge-decays}
%The only scalar field that couples to the gauge bosons is $h$ with couplings the same as the SM-Higgs:
\bea
&& W^+ W^- \longrightarrow h \qquad \qquad \frac{\sqrt{2}g^2}{2} v \nonumber\\
&& ZZ \longrightarrow h \qquad \qquad \frac{\sqrt{2}}{2}(\frac{g\cos\theta_W + g'\sin\theta_W}{2})^2 v \nonumber\\
&& H^+ H^- \longrightarrow \gamma \qquad \qquad \frac{i}{2}(g\sin\theta_W + g'\cos\theta_W)(K+K')^\mu \nonumber\\
&& H^+_{1u} H^-_{1u} \longrightarrow \gamma \qquad \qquad \frac{i}{2}(g\sin\theta_W + g'\cos\theta_W) (K+K')^\mu  \nonumber\\
&& H^+_{z} H^-_{z} \longrightarrow \gamma \qquad \qquad \frac{i}{2}(g\sin\theta_W + g'\cos\theta_W) (K+K')^\mu  \nonumber\\
&& H^+_{x} H^-_{x} \longrightarrow \gamma \qquad \qquad 2i(\frac{g\sin\theta_W + g'\cos\theta_W}{3})(K+K')^\mu \nonumber\\
&& H^+_{y} H^-_{y} \longrightarrow \gamma \qquad \qquad 2i(\frac{g\sin\theta_W + g'\cos\theta_W}{3})(K+K')^\mu \nonumber\\
&& H^\pm_{x} H^\mp_{y} \longrightarrow \gamma \qquad \qquad -i(\frac{g\sin\theta_W + g'\cos\theta_W}{3})(K+K')^\mu \nonumber\\
&& H^+ H^- \longrightarrow Z \qquad \qquad \frac{i}{2}(g\cos\theta_W - g'\sin\theta_W)(K+K')^\mu \nonumber\\
&& H^+_{1u} H^-_{1u} \longrightarrow Z \qquad \qquad \frac{i}{2}(g\cos\theta_W - g'\sin\theta_W) (K+K')^\mu  \nonumber\\
&& H^+_{z} H^-_{z} \longrightarrow Z \qquad \qquad \frac{i}{2}(g\cos\theta_W - g'\sin\theta_W) (K+K')^\mu  \nonumber\\
&& H^+_{x} H^-_{x} \longrightarrow Z \qquad \qquad 2i(\frac{g\cos\theta_W - g'\sin\theta_W}{3})(K+K')^\mu \nonumber\\
&& H^+_{y} H^-_{y} \longrightarrow Z\qquad \qquad 2i(\frac{g\cos\theta_W - g'\sin\theta_W}{3})(K+K')^\mu \nonumber\\
&& H^\pm_{x} H^\mp_{y} \longrightarrow Z \qquad \qquad -i(\frac{g\cos\theta_W - g'\sin\theta_W}{3})(K+K')^\mu \nonumber\\
&& H^\pm H \longrightarrow W^\pm \qquad \qquad \frac{ig}{2} (K+K')^\mu \nonumber\\
&& H^\pm_{1u} H^0_{1u} \longrightarrow W^\pm \qquad \qquad \frac{ig}{2} (K+K')^\mu \nonumber\\
&& H^\pm_{z} H_{z} \longrightarrow W^\pm \qquad \qquad ig (K+K')^\mu  \nonumber\\
&& H^\pm_{x} H_{x} \longrightarrow W^\pm \qquad \qquad -\frac{2ig}{3} (K+K')^\mu  \nonumber\\
&& H^\pm_{y} H_{y} \longrightarrow W^\pm\qquad \qquad -\frac{2ig}{3} (K+K')^\mu \nonumber\\
&& H^\pm_{x} H_{y} \longrightarrow W^\pm \qquad \qquad \frac{ig}{3} (K+K')^\mu \nonumber\\
&& H^\pm_{y} H_{x} \longrightarrow W^\pm \qquad \qquad \frac{ig}{3} (K+K')^\mu \nonumber\\
&& H^\pm A \longrightarrow W^\pm \qquad \qquad \frac{g}{2} (K+K')^\mu \nonumber\\
&& H^\pm_{1u} A^0_{1u} \longrightarrow W^\pm \qquad \qquad \frac{g}{2} (K+K')^\mu \nonumber\\
&& H^\pm_{z} A_{z} \longrightarrow W^\pm \qquad \qquad g (K+K')^\mu  \nonumber\\
&& H^\pm_{x} A_{x} \longrightarrow W^\pm \qquad \qquad -\frac{2g}{3} (K+K')^\mu  \nonumber\\
&& H^\pm_{y} A_{y} \longrightarrow W^\pm\qquad \qquad -\frac{2g}{3} (K+K')^\mu \nonumber\\
&& H^\pm_{x} A_{y} \longrightarrow W^\pm \qquad \qquad \frac{g}{3} (K+K')^\mu \nonumber\\
&& H^\pm_{y} A_{x} \longrightarrow W^\pm \qquad \qquad \frac{g}{3} (K+K')^\mu \nonumber\\
&& H A \longrightarrow Z \qquad \qquad \frac{1}{2}(g\cos\theta_W + g'\sin\theta_W) (K+K')^\mu \nonumber\\
&& H^0_{1u} A^0_{1u} \longrightarrow Z \qquad \qquad \frac{1}{2}(g\cos\theta_W + g'\sin\theta_W) (K+K')^\mu \nonumber\\
&& H_{z} A_{z} \longrightarrow Z \qquad \qquad (g\cos\theta_W + g'\sin\theta_W) (K+K')^\mu  \nonumber\\
&& H_{x} A_{x} \longrightarrow Z \qquad \qquad -2(\frac{g\cos\theta_W + g'\sin\theta_W}{3}) (K+K')^\mu  \nonumber\\
&& H_{y} A_{y} \longrightarrow Z \qquad \qquad -2(\frac{g\cos\theta_W + g'\sin\theta_W}{3}) (K+K')^\mu  \nonumber\\
&& H_{x} A_{y} \longrightarrow Z \qquad \qquad  (\frac{g\cos\theta_W + g'\sin\theta_W}{3}) (K+K')^\mu  \nonumber\\
&& H_{y} A_{x} \longrightarrow Z \qquad \qquad  (\frac{g\cos\theta_W + g'\sin\theta_W}{3}) (K+K')^\mu  \nonumber
\eea

%%%%%%%%%%%%%%%%%%%%%%%%%%%%%%%%%%%%%%%%%%%%%%%%%%%%%%%%%%%%%%%%%%%%%%%%%%%%%%%%

\section{Feynman rules in the I(4+2)HDM with $\tan\beta \neq 1$}\label{general}

\subsection{Scalar couplings}
\bea
&& h  \longrightarrow h h \qquad \qquad  \lambda_a v' { +2c_\beta s_\beta(2c_\beta s_\beta +c^2_\beta - s^2_\beta)\lambda_{aa}v' } \nonumber\\
&& h  \longrightarrow H H \qquad \qquad  \lambda_a v' { -2c_\beta s_\beta(2c_\beta s_\beta +c^2_\beta - s^2_\beta)\lambda_{aa}v' } \nonumber\\
&& h  \longrightarrow A A \qquad \qquad  \lambda_a v' { -2c_\beta s_\beta(2c_\beta s_\beta +c^2_\beta - s^2_\beta)\lambda_{aa}v' } \nonumber\\
&& h  \longrightarrow H^+ H^- \qquad \qquad 2 \lambda_a v' { +4c_\beta s_\beta(-2c_\beta s_\beta +c^2_\beta - s^2_\beta)\lambda_{aa}v' } \nonumber\\
&& { H  \longrightarrow hh \qquad \qquad \biggl(2 +4c_\beta s_\beta(-2 c_\beta s_\beta +c^2_\beta - s^2_\beta) \biggr) \lambda_{aa}v' } \nonumber\\
&& h  \longrightarrow  H^+_{1u} H^-_{1u} \qquad \qquad \lambda_{ai}v' \nonumber\\
&& h  \longrightarrow  H^+_{z} H^-_{z}  \qquad \qquad\lambda_{ai}v' \nonumber\\
&& h  \longrightarrow  H^+_{x} H^-_{x}  \qquad \qquad \frac{4}{3}\lambda_{ai} v' \nonumber\\
&& h  \longrightarrow  H^+_{y} H^-_{y}  \qquad \qquad \frac{4}{3}\lambda_{ai} v' \nonumber\\
&& h  \longrightarrow  H^\pm_{x} H^\mp_{y}  \qquad \qquad -\frac{2}{3}\lambda_{ai} v' \nonumber\\
&& h  \longrightarrow  H^0_{1u} H^0_{1u}  \qquad \qquad (\frac{\lambda_{ai}+\lambda'_{ai}}{2} + (1+2 c_\beta s_\beta)\lambda'_1  )v' \nonumber\\
&& h  \longrightarrow  A^0_{1u} A^0_{1u}  \qquad \qquad (\frac{\lambda_{ai}+\lambda'_{ai}}{2} - (1+2 c_\beta s_\beta)\lambda'_1  )v' \nonumber\\
&& h  \longrightarrow  H_z H_z  \qquad \qquad (\frac{\lambda_{ai} + \lambda'_{ai}}{2}+ 3(1+2c_\beta s_\beta)\lambda_1  ) v' \nonumber\\
&& h  \longrightarrow  H_x H_x  \qquad \qquad (\frac{2\lambda_{ai}+2\lambda'_{ai}}{3} ) v' \nonumber\\
&& h  \longrightarrow  H_y H_y  \qquad \qquad (\frac{2\lambda_{ai}+2\lambda'_{ai}}{3} ) v' \nonumber\\
&& h  \longrightarrow  H_x H_y  \qquad \qquad -(\frac{2\lambda_{ai}+2\lambda'_{ai}}{3} ) v' \nonumber\\
&& h  \longrightarrow  A_z A_z  \qquad \qquad (\frac{\lambda_{ai} + \lambda'_{ai}}{2} -3(1+2c_\beta s_\beta)\lambda_1   ) v' \nonumber\\
&& h  \longrightarrow  A_x A_x  \qquad \qquad (\frac{2\lambda_{ai}+2\lambda'_{ai}}{3} ) v' \nonumber\\
&& h  \longrightarrow  A_y A_y  \qquad \qquad (\frac{2\lambda_{ai}+2\lambda'_{ai}}{3} ) v' \nonumber\\
&& h  \longrightarrow  A_x A_y  \qquad \qquad -(\frac{2\lambda_{ai}+2\lambda'_{ai}}{3} ) v' \nonumber\\
&& H   \longrightarrow  H_{z} H_{z} \qquad \qquad     3 \lambda_1(c^2_\beta - s^2_\beta) v' \nonumber\\
&& H   \longrightarrow  A_{z} A_{z} \qquad \qquad    -3 \lambda_1(c^2_\beta - s^2_\beta) v' \nonumber\\
&& H^\pm \longrightarrow  H^\pm_{z} H_{z} \qquad \qquad 3 \lambda_1(c^2_\beta - s^2_\beta) v' \nonumber\\
&& H^\pm \longrightarrow  H^\pm_{z} A_{z} \qquad \qquad \mp 3 i \lambda_1(c^2_\beta - s^2_\beta) v' \nonumber\\
&& A  \longrightarrow  H_z A_z  \qquad \qquad   2 \lambda_1(c^2_\beta - s^2_\beta) v' \nonumber\\
&& A  \longrightarrow  H_x A_x  \qquad \qquad    \frac{4}{3}\lambda_1(c^2_\beta - s^2_\beta) v' \nonumber\\
&& A  \longrightarrow  H_y A_y  \qquad \qquad    \frac{4}{3}\lambda_1(c^2_\beta - s^2_\beta) v' \nonumber\\
&& A  \longrightarrow  H_x A_y  \qquad \qquad   -\frac{2}{3}\lambda_1(c^2_\beta - s^2_\beta) v' \nonumber\\
&& A  \longrightarrow  H_y A_x  \qquad \qquad   -\frac{2}{3}\lambda_1(c^2_\beta - s^2_\beta) v' \nonumber\\ 
&& A    \longrightarrow  H^0_{1u} A^0_{1u}   \qquad \qquad   2\lambda'_1 (c^2_\beta - s^2_\beta)  v' \nonumber\\
&& H \longrightarrow  H^0_{1u} H^0_{1u}   \qquad \qquad    \lambda'_1 (c^2_\beta - s^2_\beta)  v' \nonumber\\
&& H    \longrightarrow  A^0_{1u} A^0_{1u}   \qquad \qquad   -\lambda'_1 (c^2_\beta - s^2_\beta)  v' \nonumber\\
&& H^\pm \longrightarrow  H^\pm_{1u} H^0_{1u} \qquad \qquad    \lambda'_1 (c^2_\beta - s^2_\beta)  v' \nonumber\\
&& H^\pm \longrightarrow  H^\pm_{1u} A^0_{1u} \qquad \qquad \mp i\lambda'_1 (c^2_\beta - s^2_\beta)  v' \nonumber
\eea

\subsection{Gauge couplings}

\bea
&& W^+ W^- \longrightarrow h     \qquad \qquad \frac{g^2}{2} v' \nonumber\\
&& Z Z     \longrightarrow h     \qquad \qquad \frac{\sqrt{2}}{2}\biggl(\frac{g\cos\theta_W + g'\sin\theta_W}{2} \biggr)^2   v' \nonumber\\
&& H^+ H^- \longrightarrow \gamma \qquad \qquad \frac{i}{2}(g\sin\theta_W + g'\cos\theta_W)(K+K')^\mu \nonumber\\
&& H^+_{1u} H^-_{1u} \longrightarrow \gamma \qquad \qquad \frac{i}{2}(g\sin\theta_W + g'\cos\theta_W) (K+K')^\mu  \nonumber\\
&& H^+_{z} H^-_{z} \longrightarrow \gamma \qquad \qquad \frac{i}{2}(g\sin\theta_W + g'\cos\theta_W) (K+K')^\mu  \nonumber\\
&& H^+_{x} H^-_{x} \longrightarrow \gamma \qquad \qquad 2i(\frac{g\sin\theta_W + g'\cos\theta_W}{3})(K+K')^\mu \nonumber\\
&& H^+_{y} H^-_{y} \longrightarrow \gamma \qquad \qquad 2i(\frac{g\sin\theta_W + g'\cos\theta_W}{3})(K+K')^\mu \nonumber\\
&& H^\pm_{x} H^\mp_{y} \longrightarrow \gamma \qquad \qquad -i(\frac{g\sin\theta_W + g'\cos\theta_W}{3})(K+K')^\mu \nonumber\\
&& H^+ H^- \longrightarrow Z \qquad \qquad \frac{i}{2}(g\cos\theta_W - g'\sin\theta_W)(K+K')^\mu \nonumber\\
&& H^+_{1u} H^-_{1u} \longrightarrow Z \qquad \qquad \frac{i}{2}(g\cos\theta_W - g'\sin\theta_W) (K+K')^\mu  \nonumber\\
&& H^+_{z} H^-_{z} \longrightarrow Z \qquad \qquad \frac{i}{2}(g\cos\theta_W - g'\sin\theta_W) (K+K')^\mu  \nonumber\\
&& H^+_{x} H^-_{x} \longrightarrow Z \qquad \qquad 2i(\frac{g\cos\theta_W - g'\sin\theta_W}{3})(K+K')^\mu \nonumber\\
&& H^+_{y} H^-_{y} \longrightarrow Z\qquad \qquad 2i(\frac{g\cos\theta_W - g'\sin\theta_W}{3})(K+K')^\mu \nonumber\\
&& H^\pm_{x} H^\mp_{y} \longrightarrow Z \qquad \qquad -i(\frac{g\cos\theta_W - g'\sin\theta_W}{3})(K+K')^\mu \nonumber\\
&& H^\pm H \longrightarrow W^\pm \qquad \qquad \frac{ig}{2}(K+K')^\mu \nonumber\\
&& H^\pm_{1u} H^0_{1u} \longrightarrow W^\pm \qquad \qquad \frac{ig}{2}( (K+K')^\mu \nonumber\\
&& H^\pm_{z} H_{z} \longrightarrow W^\pm \qquad \qquad ig (K+K')^\mu  \nonumber\\
&& H^\pm_{x} H_{x} \longrightarrow W^\pm \qquad \qquad -\frac{2ig}{3} (K+K')^\mu  \nonumber\\
&& H^\pm_{y} H_{y} \longrightarrow W^\pm\qquad \qquad -\frac{2ig}{3} (K+K')^\mu \nonumber\\
&& H^\pm_{x} H_{y} \longrightarrow W^\pm \qquad \qquad \frac{ig}{3} (K+K')^\mu \nonumber\\
&& H^\pm_{y} H_{x} \longrightarrow W^\pm \qquad \qquad \frac{ig}{3} (K+K')^\mu \nonumber\\
&& H^\pm A \longrightarrow W^\pm             \qquad \qquad \frac{g}{2}( (K+K')^\mu \nonumber\\
&& H^\pm_{1u} A^0_{1u} \longrightarrow W^\pm \qquad \qquad \frac{g}{2} (K+K')^\mu \nonumber\\
&& H^\pm_{z} A_{z} \longrightarrow W^\pm     \qquad \qquad g (K+K')^\mu  \nonumber\\
&& H^\pm_{x} A_{x} \longrightarrow W^\pm     \qquad \qquad -\frac{2g}{3} (K+K')^\mu  \nonumber\\
&& H^\pm_{y} A_{y} \longrightarrow W^\pm     \qquad \qquad -\frac{2g}{3} (K+K')^\mu \nonumber\\
&& H^\pm_{x} A_{y} \longrightarrow W^\pm     \qquad \qquad \frac{g}{3} (K+K')^\mu \nonumber\\
&& H^\pm_{y} A_{x} \longrightarrow W^\pm     \qquad \qquad \frac{g}{3} (K+K')^\mu \nonumber\\
&& H A \longrightarrow Z \qquad \qquad \frac{g}{2}(g\cos\theta_W + g'\sin\theta_W) (K+K')^\mu \nonumber\\
&& H^0_{1u} A^0_{1u} \longrightarrow Z \qquad \qquad \frac{g}{2}(g\cos\theta_W + g'\sin\theta_W) (K+K')^\mu \nonumber\\
&& H_{z} A_{z} \longrightarrow Z \qquad \qquad (g\cos\theta_W + g'\sin\theta_W) (K+K')^\mu  \nonumber\\
&& H_{x} A_{x} \longrightarrow Z \qquad \qquad -2(\frac{g\cos\theta_W + g'\sin\theta_W}{3}) (K+K')^\mu  \nonumber\\
&& H_{y} A_{y} \longrightarrow Z \qquad \qquad -2(\frac{g\cos\theta_W + g'\sin\theta_W}{3}) (K+K')^\mu  \nonumber\\
&& H_{x} A_{y} \longrightarrow Z \qquad \qquad (\frac{g\cos\theta_W + g'\sin\theta_W}{3}) (K+K')^\mu  \nonumber\\
&& H_{y} A_{x} \longrightarrow  Z \qquad \qquad (\frac{g\cos\theta_W + g'\sin\theta_W}{3}) (K+K')^\mu  \nonumber
\eea

\section{The Yukawa sector in the I(4+2)HDM}
The two active doublets $H_{3u}$ and $H_{3d}$ are the only doublets which couple to fermions. To avoid FCNCs we require the up-type and down-type fermions to couple to $H_{3u}$ and $H_{3d}$ respectively, which is the standard method in Type-II-2HDM. Therefore, the Yukawa couplings in this model are similar to that of Type-II-2HDM\footnote{Our notation matches the notation used in \cite{Report} with the following relations:
\bea 
\mu^2_5 \equiv m^2_{11}, \quad \mu^2_6 \equiv m^2_{22}, \quad \mu'^2_a \equiv -2 m^2_{12} \nonumber\\
2\lambda_a \equiv \lambda_1 = \lambda_2 = \lambda_3 , \quad 2\lambda_{aa} \equiv \lambda_4 = \lambda_5 \nonumber
\eea}\cite{Report}:
\bea
{\mathcal{L}}_Y&=& - \sum_{f=u,d,l} \frac{m_f}{v} \biggl( (\xi^f_h)\overline{f}f(h) + (\xi^f_H)\overline{f}f(H) - i (\xi^f_A)\overline{f}f(A) \biggr) \\
&&- \left[ \frac{\sqrt{2}V_{CKM}}{2} \overline{u} \biggl( m_u (\xi^u_A) P_L  + m_d (\xi^d_A) P_R
\biggr) d H^+ + \frac{\sqrt{2}m_l}{v} (\xi^l_A) \overline{\nu_L} l_R H^+ + h.c. \right] \nonumber
\eea
where $(m_f/v)$ is the SM coupling of the fermion $f$ to the Higgs boson and $P_{L/R}=(1 \pm \gamma_5)/2$.

\begin{itemize}
\item
In the $\tan\beta = 1$ case; 
$$\sin\beta=\cos\beta=\frac{\sqrt{2}}{2}$$
and 
$$\sin\alpha=-\cos\alpha=-\frac{\sqrt{2}}{2}$$ 
providing the conditions $-{\mu'_a}^2>0$ and $-{\mu'_a}^2 >2\lambda_a$ are satisfied.

Therefore, the Type-II-2HDM-like Yukawa couplings of $u,d,l$ to the neutral Higgs bosons $h, H, A$ are:
\bea
&& \xi^u_h= \frac{\cos\alpha}{\sin\beta}=1, \qquad 
   \xi^d_h= -\frac{\sin\alpha}{\cos\beta}=1, \qquad 
   \xi^l_h= -\frac{\sin\alpha}{\cos\beta}=1 \nonumber\\
&& \xi^u_H= \frac{\sin\alpha}{\sin\beta}=-1, \qquad 
   \xi^d_H= \frac{\cos\alpha}{\cos\beta}=1, \qquad 
   \xi^l_H= \frac{\cos\alpha}{\cos\beta}=1 \\
&& \xi^u_A= \cot\beta=1, \qquad 
   \xi^d_A= \tan\beta=1, \qquad 
   \xi^l_A= \tan\beta=1 \nonumber
\eea

\item
In the $\tan\beta \neq 1$ case, angle $\alpha$ is calculated from the rotation of the neutral CP-even mass matrix;
\be 
M = \left( \begin{array}{cc}
\frac{\mu^2_5}{2}+\frac{\lambda_a(3v^2_u+v^2_d)}{2}+ \lambda_{aa} v^2_d \quad & \quad \frac{\mu'^2_a}{2}+2\lambda_a v_u v_d + 4\lambda_{aa} v_u v_d  \\ [2mm]
\frac{\mu'^2_a}{2}+2\lambda_a v_u v_d + 4\lambda_{aa} v_u v_d   \quad & \quad  \frac{\mu^2_6}{2}+\frac{\lambda_a(v^2_u+3v^2_d)}{2} + \lambda_{aa} v^2_u 
\end{array} \right) 
\ee
which leads to:
\be 
\tan(2\alpha) = \frac{2M_{12}}{M_{11}-M_{22}} = \quad \frac{2\mu'^2_a + 4(\lambda_a +2\lambda_{aa}) v'^2 \sin(2\beta)}{(\mu^2_5 - \mu^2_6) + 2(\lambda_a -\lambda_{aa}) v'^2 \cos(2\beta)}.
\ee

Therefore, the Yukawa couplings of $u,d,l$ to the neutral Higgs bosons $h, H, A$ are:
\bea
&& \xi^u_h= \frac{\cos\alpha}{\sin\beta}, \qquad 
   \xi^d_h= -\frac{\sin\alpha}{\cos\beta}, \qquad 
   \xi^l_h= -\frac{\sin\alpha}{\cos\beta} \nonumber\\
&& \xi^u_H= \frac{\sin\alpha}{\sin\beta}, \qquad 
   \xi^d_H= \frac{\cos\alpha}{\cos\beta}, \qquad 
   \xi^l_H= \frac{\cos\alpha}{\cos\beta} \\
&& \xi^u_A= \cot\beta, \qquad 
   \xi^d_A= \tan\beta, \qquad 
   \xi^l_A= \tan\beta \nonumber
\eea

\end{itemize}

\end{document}